%% file: ms.tex
\documentclass[1p]{elsarticle}

\usepackage{amssymb, amsmath, amsfonts, amsthm, wasysym, mathtools} 
\usepackage{algorithm,algpseudocode, setspace,placeins}
\usepackage{epsfig,subcaption} 
\usepackage{fullpage, url, color} 

\usepackage{longtable,booktabs}
\usepackage{bm}
\usepackage[textsize=tiny,disable]{todonotes}
\usepackage{float}

\usepackage{xr-hyper}
\usepackage{comment}

\usepackage{pdfpages}

\usepackage{lineno}

 \usepackage{varioref}
 \usepackage[colorlinks=true,
 	    linkcolor=blue,
 	    urlcolor=blue,
 	    citecolor=blue,
 	    final,
 	    hypertexnames=false,
 	    linktoc=all]{hyperref}
\usepackage[noabbrev]{cleveref}
\crefname{appendix}{Appendix}{Appendices}
\crefname{system}{system}{systems}
\Crefname{system}{System}{Systems}
\creflabelformat{system}{#2(#1)#3}
\crefname{step}{step}{steps}
\Crefname{step}{Step}{Steps}

\allowdisplaybreaks

\usepackage{mydefs}

\begin{document}

\begin{frontmatter}
\title{A pseudospectral method for direct numerical simulation of
  low-Mach, variable-density, turbulent flows}

\author[sandia]{Bryan W.~Reuter\corref{bryan}}
\ead{bwreute@sandia.gov}

\author[pecos]{Todd A.~Oliver}
\ead{oliver@oden.utexas.edu}

\author[pecos,meche]{Robert D.~Moser}
\ead{rmoser@oden.utexas.edu}

\address[sandia]{Sandia National Laboratories}
\address[pecos]{Center for Predictive Engineering and Computational Sciences,\\
        Oden Institute for Computational Engineering and Sciences, The University of Texas-Austin}
\address[meche]{Department of Mechanical Engineering, The University of Texas-Austin}

\cortext[bryan]{Corresponding author}

\begin{abstract}
A novel algorithm for the direct numerical simulation of the
variable-density, low-Mach Navier-Stokes equations extending the
method of Kim, Moin, and Moser \cite{kim1987turbulence} for
incompressible flow is presented here.  A Fourier representation is
employed in the two homogeneous spatial directions and a number of
discretizations can be used in the inhomogeneous direction.
The momentum is decomposed into
divergence- and curl-free portions which allows the momentum equations
to be rewritten, removing the need to solve for the
pressure.  The temporal discretization is based on an explicit,
segregated Runge-Kutta method and the scalar equations are
reformulated to directly address the redundancy of the equation of
state and the mass conservation equation.  An efficient, matrix-free,
iterative solution of the resulting equations allows for second-order
accuracy in time and numerical stability for large density ratios,
which is demonstrated for ratios up to $\sim 25.7$.
\end{abstract}



\end{frontmatter}

\listoftodos

\section{Introduction}
\label{sec:intro}
\input{intro.tex}

\section{Governing equations}
\label{sec:lmns}
\input{lmns.tex}

\section{Temporal discretization}
\label{sec:temp_disc}
\input{temp_disc.tex}

\section{Discretizing the variable-density, low-Mach Navier-Stokes equations}
\label{sec:algo}
\input{algo.tex}

\section{Results}
\label{sec:results}
\input{results.tex}

\section{Conclusions}
\label{sec:conclusion}
\input{conclusion.tex}

\section{Acknowledgments}
\noindent Sandia National Laboratories is a multimission laboratory managed and operated by National Technology \& Engineering Solutions of Sandia, LLC, a wholly owned subsidiary of Honeywell International Inc., for the U.S. Department of Energy’s National Nuclear Security Administration under contract DE-NA0003525.
This paper describes objective technical results and analysis. Any subjective views or opinions that might be expressed in the paper do not necessarily represent the views of the U.S. Department of Energy or the United States Government.

\clearpage 
\appendix

\section{Obtaining the BDF-like approximations}
\label{app:coeffs}
\input{coeffs.tex}
\section{Matrices used in linear stability analysis}
\label{app:lstab}
\input{lstab.tex}
\section{Spatial discretization}
\label{app:spat_disc}
\input{spat_disc.tex}

\section{Mesh convergence and simulation specifics for Rayleigh-Taylor problem}
\label{app:mesh_conv}
\input{mesh_conv.tex}

\clearpage 
\bibliographystyle{elsarticle-harv}
\bibliography{references}
\end{document}

%% file: intro.tex
In this paper, we seek to enable the direct numerical simulation (DNS) of low-speed turbulent flows with significant density variations.
Such flows are common in engineering and nature, such as in turbulent combustion, nonreacting mixing problems, the atmosphere, and the oceans.
In the low-Mach-number limit, the pressure fluctuations are decoupled from the density fluctuations. 
The pressure can then be decomposed into a uniform background, thermodynamic pressure and a mechanical, or dynamic, pressure.
The mechanical pressure fluctuates in space and time and acts to enforce mass conservation as in incompressible flow.
Unlike constant-density incompressible flow, in variable-density flow, there is a time derivative term in the mass conservation equation which presents additional numerical challenges, particularly with large density variation.

It is natural to adapt algorithms designed for constant-density, incompressible flows to the low-Mach-number, variable-density case due to the similar role of the mechanical pressure.
Fully compressible algorithms perform poorly at low Mach numbers, since the Navier-Stokes equations are formally singular as $\text{Ma} \rightarrow 0$.
Furthermore, compressible solvers are built to handle often severe stability constraints imposed by acoustic timescales which are removed by the low-Mach assumption.
Therefore, most numerical methods for solving the low-Mach equations are based on traditional incompressible fractional-step, projection approaches \cite{Knikker2011,Nicoud2000}.
These formulations differ based on a choice to enforce either the divergence of the momentum or the divergence of the velocity, resulting in a constant-coefficient or variable-coefficient pressure Poisson equation, respectively.
In both cases, the time dependence of the constraint necessitates particular care in enforcing the constraint in the context of the time discretization scheme.
\subsection{Treatment of the mechanical pressure in fractional step methods}
\label{subsec:press}
Fundamental to the low-Mach Navier-Stokes equations is the time-dependent divergence constraint on the momentum and the role of the mechanical pressure in enforcing that constraint.
Fractional-step approaches for incompressible flows solve for an intermediate velocity or momentum field by neglecting (``pressure-free" methods \cite{brown2001accurate}) or lagging (``incremental'' methods \cite{GUERMOND20066011}) the pressure in the momentum equations and then projecting the result onto a divergence-free basis through the solution of a Poisson equation derived from the momentum equations.
Since the momentum field is not divergence free in variable-density flows, standard incompressible projection methods must be modified.
Within the class of fractional-step methods for the low-Mach equations, two formulations are commonly employed.
One option is to take the divergence of the momentum equations, giving a constant-coefficient Poisson equation for the pressure.
Upon temporal discretization, the divergence of the momentum is related to the time derivative of the density through the mass conservation equation.
The second option involves manipulating the advective form of the momentum equations after dividing through by the density to give a variable-coefficient Poisson equation. 
The velocity divergence is then related to the material derivative of the density by reformulating the mass conservation equation.

The first approach has the advantage of being much easier to solve, but the evaluation of the density time derivative can lead to instability, especially when density ratios are larger than three \cite{Knikker2011,Rauwoens2009,Motheau2016}.
To help alleviate this issue, predictor-corrector schemes or implicit formulations have shown some success \cite{Najm1998,Wall2002,Jang2007,domino_predicting_2021}.
Predictor-corrector methods are attractive in scenarios where only explicit or segregated time-stepping is practical (e.g.~when spectral methods are employed) \cite{Jang2007}.
The second approach \cite{Nicoud2000,bell2005amr} does allow for higher density ratios, but cannot take advantage of the wide array of efficient solution algorithms for the pressure Poisson equation in incompressible flows.
The solution of the variable-coefficient Poisson equation generally requires an iterative scheme, can be an order of magnitude slower, and convergence can be hampered by large density ratios \cite{Dodd2014}.
As always with such projection methods, both approaches must specify boundary conditions for the pressure which are not formally specified \textit{a priori} and can introduce inconsistencies with the velocity field \cite{Orszag1986}.
A third option, in which the pressure is eliminated entirely from the dynamic equations, is used in this work.
This is advantageous because only a cheap, constant-coefficient Poisson solve for the momentum, rather than the pressure, is required. 
Furthermore, all boundary conditions are well defined.
\subsection{The redundant nature of the low-Mach equations and its impact on numerical stability} \label{subsec:redund}
Formally, the low-Mach Navier-Stokes equations are a partial differential-algebraic system (coupled partial differential equations along with the conservation of mass constraint and an equation of state).
However, there is a redundancy in the equations since the density must simultaneously obey the equation of state and the mass conservation equation.
Inconsistencies between the scalar fields, the equation of state, and the momentum field have been identified as numerically destabilizing, as kinetic energy can be incorrectly injected into the system nonlocally through the divergence constraint on the momentum \cite{Pierce2001}.
Work by Shunn and Ham \cite{shunn2006consistent} suggests that scenarios where the equation of state is sufficiently nonlinear can introduce prohibitive resolution requirements on the density even when the scalar fields are well resolved. 
In turn, these underresolved features can produce a nonphysical velocity field.
More generally, instability can occur due to inconsistencies that arise when using a segregated time advancement scheme.
Knikker \cite{Knikker2011} notes that the structure of the low-Mach equations leads to redundancies that make it impossible to both advance them in conservative form and satisfy the equation of state without resorting to a temporally implicit scheme, which can be cost prohibitive. 
Instead predictor-corrector approaches are common, since they presumably can lessen, but not eliminate, the degree of the discrepancies and provide additional stability while remaining computationally cheaper.
These methods are susceptible to instabilities when density gradients are large \cite{Motheau2016}.
Herein, we demonstrate how the redundancy can be eliminated and the potential for instability due to inconsistencies removed.
Further, to facilitate efficient solution in turbulence simulation, a novel time integration scheme is developed and made robust to large density variations.
\subsection{Outline of the present approach}
%
To address challenges surrounding the mechanical pressure and stability in the context of direct numerical simulation, where highly accurate numerics are necessary and computational demands are extreme, a novel algorithm has been developed.
A DNS solves the Navier-Stokes equations for the full, instantaneous fields without introducing any models for the turbulence, making it an invaluable tool for model development and answering fundamental questions in fluid mechanics.
However, the demands of resolving all lengthscales and timescales present are high and DNS is currently limited to canonical configurations with two or three statistically homogeneous directions and must take advantage of numerical techniques which are massively parallelizable to retain tractability.
The formulation presented here is designed for scenarios with two statistically homogeneous directions that can be simulated with periodic boundary conditions, enabling the use of Fourier spectral methods, and one inhomogeneous direction.
It is an extension of the scheme of Kim, Moin, and Moser (KMM) \cite{kim1987turbulence} developed for incompressible flows.
In their scheme, equations for one component of the vorticity and the Laplacian of the velocity are advanced, eliminating the need to solve for the pressure.
Herein the momentum is decomposed into divergence-free and curl-free parts.
The divergence-free momentum plays a role that is similar to the incompressible velocity in the original KMM algorithm, and the curl-free momentum is determined from the mass conservation equation.
A similar decomposition was used in the work of Almagro\etal \cite{almagro2017numerical}.
Their scheme, however, is only first-order accurate in time despite using three Runge-Kutta (RK) stages and assumes constant fluid viscosity, thermal conductivity, and specific heat.
The algorithm presented here relaxes these restrictive assumptions, achieves a higher-order temporal accuracy in two stages while retaining robustness for flows with high density ratios, and \textit{a priori} guarantees discrete conservation of mass.

This paper is organized as follows.
Firstly, \cref{sec:lmns} introduces the governing equations, the momentum decomposition approach, and explicitly identifies the source of numerical instability targeted by the new time discretization strategy.
\Cref{sec:temp_disc} describes the temporal discretization, its second-order convergence, and its numerical stability.
\Cref{sec:algo} lays out the DNS algorithm and demonstrates its efficacy for flows with large density ratios via a single-mode Rayleigh-Taylor test case.

%% file: lmns.tex
With the understanding that the method presented here extends to the general case of $N_s$ species and an energy equation, the following will be limited to a formulation in which a single conserved scalar $z$ is used to characterize the thermochemical state.
The low-Mach-number Navier-Stokes equations representing conservation of mass, momentum, and $z$ for a viscous fluid are, along with the equation of state,
\begin{equation} \label[system]{eqn:LMNS_system}
\begin{aligned}
\pp{\rho}{t} + &\pp{\rho u_j}{x_j} = 0 \\
\pp{\rho u_i}{t} + \pp{\rho u_i u_j}{x_j} &= - \pp{p}{x_i} + \pp{\tau_{ij}}{x_j} \\
\pp{\rho z}{t} + \pp{}{x_j} \left ( \rho u_j z \right ) &= \pp{}{x_j} \left ( \rho \mc{D}_z \pp{z}{x_j} \right ) \\
\rho = &\; f(z) 
\end{aligned}
\end{equation}
where $u_i$ is the fluid velocity, $\rho$ the density, $p$ the mechanical pressure, $\tau_{ij}$ the viscous stress tensor, and $\rho \mc{D}_z$ is the effective diffusivity of $z$.
For a Newtonian fluid the stress tensor is
\begin{equation}
\tau_{ij} = -\dfrac{2}{3} \mu \pp{u_k}{x_k} \delta_{ij} + 
			\mu \left ( \pp{u_i}{x_j} + \pp{u_j}{x_i} \right )
\end{equation}
where $\mu$ is the fluid dynamic viscosity and $\delta_{ij}$ is the Kronecker delta.
The passive scalar, $z$, is assumed to exhibit Fickian diffusive transport and, for example, in the simplified case of inert, binary mixing or reduced order descriptions of nonpremixed chemical reactions, $z$ represents the local fluid composition.
The transport coefficients $\mu = g(z)$ and $\mc{D}_z = h(z)$ are assumed to be known functions of $z$, as is the density through the equation of state $\rho = f(z)$.

At this point we introduce the Helmholtz decomposition of the momentum $\bm{m} = \rho \bm{u}$:
\begin{equation} \label{eqn:mom_decomp}
\bm{m} = \bm{m}^d + \bm{m}^c
\end{equation}
where $\bm{m}^d$ is divergence free and $\bm{m}^c$ is curl free, so that $\bm{m}^c = \grad{\psi}$ for some scalar potential $\psi$.
With this decomposition, the conservation of mass equation is
\begin{equation} \label{eqn:COMass}
\pp{\rho}{t} = -\div{\bm{m}^c} = -\lap{\psi} \, ,
\end{equation}
which represents a constraint on the curl-free momentum.
The momentum equations can then be thought of as dynamic equations for $\bm{m}^d$ once rewritten as
\begin{equation} \label{eqn:COMom}
\pp{m^d_i}{t} + \pp{\rho u_i u_j}{x_j} = - \pp{\zeta}{x_i} + \pp{\tau_{ij}}{x_j}
\end{equation}
where $\zeta = p + \partial \psi / \partial t$.
\subsection{Reframing the scalar transport equation} \label{subsec:L}
The variable-density Navier-Stokes equations are a system of partial differential equations along with the equation of state that makes the mass conservation equation redundant.
The equation of state $\rho = f(z)$, and the mass conservation equation imply
\begin{equation} \label{eqn:COM_with_psi}
\pp{\rho}{t} = \dd{f}{z}\pp{z}{t} = -\div{\bm{m}^c} = -\lap{\psi} \, .
\end{equation}
Formally solving \cref{eqn:COM_with_psi} for $\psi$ and applying the gradient yields
\begin{equation} \label{eqn:lap_mc_const}
\bm{m}^c = -\grad{\left(\nabla^2\right)^{-1} \left[\dd{f}{z}\pp{z}{t}\right]}
\end{equation}
where $\left(\nabla^2\right)^{-1}$ is the inverse Laplacian, defined with an appropriate set of boundary conditions.
When time advancing the equations, the conserved scalar $z$ and the momentum will not satisfy \cref{eqn:lap_mc_const} in general.
As noted in \cref{subsec:redund}, such inconsistencies are destabilizing, especially when large density variations are present.

To remove the potential for numerical instability, the $z$ equation can be reformulated as follows:
let
\begin{gather}
\mc{L}(z)\left\{\cdot\right\} \equiv - \dfrac{1}{\rho} \grad{z} \cdot \grad{\left(\nabla^2\right)^{-1} \left[\dd{f}{z}\{\cdot\}\right]} \, , \\
\mc{R}_z \equiv - \dfrac{1}{\rho} \bm{m}^d \cdot \grad{z} + \dfrac{1}{\rho} \div{\left(\rho\mc{D}_z \grad{z}\right)} \, ,
\end{gather}
noting that $\mc{L}$ is linear in $\partial z / \partial t$ and depends on $z$.
Then the scalar transport equation can be rewritten as
\begin{equation} \label{eqn:zeqn_imp}
\pp{z}{t} = \mc{L}(z)\pp{z}{t} + \mc{R}_z \, .
\end{equation}
%
%
This makes explicit the dependence of the right-hand side on $\partial z / \partial t$ due to the $\bm{m}^c$ portion of the convective term.
%
%
%
%
If $\mc{I} - \mc{L}$ is nonsingular, \cref{eqn:zeqn_imp} can be
rewritten in explicit form:
\begin{equation} \label{eqn:zeqn_exp}
\pp{z}{t} = \left\{\mc{I} - \mc{L}(z)\right\}^{-1} \mc{R}_z \, .
\end{equation}
Evaluating $\partial z / \partial t$ in this way ensures consistency between the equation of state, the mass conservation equation, and the scalar transport equation, avoiding potential instability because the redundant equations have been eliminated.
\Cref{eqn:zeqn_exp} can be time advanced with standard methods to update $z$ and then the right-hand side used in \cref{eqn:lap_mc_const} to obtain a consistent $\bm{m}^c$.
However, in the turbulence DNS targeted here, inverting $\mc{I} - \mc{L}$ directly is not practical.
Instead we introduce an incomplete iterative, matrix-free solution method (\cref{sec:temp_disc}) which allows the $z$ equation to be time-advanced efficiently and stably.
%

%% file: temp_disc.tex
Fourier spectral methods rely on a wavenumber-by-wavenumber decoupling to make computation tractable.
In the variable-density case, few operators in the governing equations are linear, so explicit time integration or linearly-implicit schemes that avoid prohibitively expensive nonlinear solves are attractive.
This work focuses on the DNS of flows in unbounded domains where the timestep required for accuracy is similar to the timestep required for stability when using an explicit method \cite{moin1998direct}.
For this reason, it is based on the explicit, second-order Runge-Kutta scheme.
Explicit time discretization also allows the momentum and $z$ equations to be advanced independently.
However, treatment of the $z$ equation is complicated by the mass conservation constraint, which is particularly challenging to enforce for large density ratios.

A new, robust numerical formulation for solving the $z$ equation is introduced below that employs a fixed-point iteration which, upon convergence, inverts $\mc{I} - \mc{L}$ and mitigates the stability problems discussed in \cref{subsec:L}.
Recognizing that full convergence will not be necessary for stability, the method is designed so that $z$ will be temporally second order, consistent with the temporal discretization of the momentum equations, regardless of the number of iterations.
Hence, the number of iterations is selected to reduce the destabilizing inconsistencies between the equation of state, the mass conservation equation, and the scalar transport equation to an acceptable level.
This allows for stability while minimizing cost.

As the scheme includes both a fixed-point problem and temporal discretization, we begin by considering a linear fixed-point problem 
\begin{equation} \label{eqn:linear_fixedpoint}
y = \mc{A} y + b
\end{equation}
and, under the assumptions that inverting $\mc{I} - \mc{A}$ directly is not feasible and that the spectral radius of $\mc{A}$ is less than one, demonstrate how \cref{eqn:linear_fixedpoint} can be solved approximately such that
\begin{enumerate}
\item the number of iterations does not affect the order of accuracy of the solution;
\item the iterations converge to the exact solution $y = \left\{ \mc{I} - \mc{A} \right\}^{-1} b$.
\end{enumerate}
Recognizing that the $z$ equation (\cref{eqn:zeqn_imp}) can be solved
in this way to obtain evaluations of $\partial z / \partial t$, we then show how this technique can be integrated into a RK2 time advancement of $z$ without damaging second-order convergence.
\subsection{Incomplete solution of the fixed-point problem} \label{subsec:incomp}
Assuming the direct solution of \cref{eqn:linear_fixedpoint} is not practical, a straightforward solution is then to take an iterative approach:
\begin{equation} \label{eqn:y_k}
y^k = \mc{A} y^{k-1} + b
\end{equation}
where $y^k$ is the $k^{th}$ iterate.
\Cref{eqn:y_k} can be rewritten in terms of an initial guess $y^0$:
\begin{equation} \label{eqn:y_k_init_guess}
y^k = \mc{A}^{k} y^0 + \sum_{j=0}^{k-1} \mc{A}^j b \, .
\end{equation}
Since $\rho(\mc{A}) < 1$, $\mc{A}^k y^0 \xrightarrow[k\rightarrow\infty]{} 0$ and $\left\{\mc{I}-\mc{A}\right\}^{-1} = \mc{I} + \mc{A} + \mc{A}^2 + \cdots \, = \sum_{j=0}^{\infty} \mc{A}^j$.
Hence, in the limit of large $k$ the solution to the fixed-point problem approaches the exact solution $y = \left\{\mc{I}-\mc{A}\right\}^{-1}b$.

Now further assume the initial guess is such that $y^0 = y + \delta y$
where $y$ is the exact solution and $\delta y$ is an error which goes
to zero asymptotically with an order parameter $\epsilon$, so that $\delta y = \epsilon c$ for $c \sim \mc{O}(1)$.
\Cref{eqn:y_k_init_guess} becomes
\begin{equation} \label{eqn:y_k_delta}
\begin{aligned}
y^k &= \mc{A}^{k} ( y + \delta y ) + \sum_{j=0}^{k-1} \mc{A}^j b = \mc{A}^{k} y + \sum_{j=0}^{k-1} \mc{A}^j b + \mc{A}^k \delta y \, .
\end{aligned}
\end{equation}
Since $y = \mc{A} y + b$, the first two terms simplify due to the identity $\mc{A}^{j-1} b = \mc{A}^{j-1} y - \mc{A}^j y$.
The $k^{th}$ iterate then satisfies $y^k = y + \mc{A}^k \delta y$, which indicates the error in the $k^{th}$ iterate is $\mc{A}^k \delta y$.
It follows that
\begin{equation} \label{eqn:y_k_orderep}
y^k = y + \epsilon \mc{A}^k c \, .
\end{equation}
Hence, the error in the $k^{th}$ iterate is also $\mc{O}(\epsilon)$ --
the order of accuracy of the solution is independent of the number of
iterations.  This is helpful when applications of $\mc{A}$ are
expensive so that converging the iterations is prohibitive,
but useful approximations of $y$ are available to use as a starting guess.
\subsection{Advancing the scalar transport equation} \label{subsec:fixedpoint_for_dns}
For the rest of this section assume the equations are semi-discrete (discretized in space) so that they are systems of ODEs.
It will be implicitly understood that $\mc{L}$, for example, now represents the spatially discretized operator previously introduced.
The time advancement of $z$ from time level $n$ to $n+1$ by the RK2 scheme requires evaluations $\left(dz / dt\right)^{n}$ and $\left(dz / dt\right)^{'}$ which are accurate up to $\mc{O}\left(\dt^2\right)$.
The prime notation indicates the intermediate stage of RK2 which is associated with the time $t_{n+1/2} = t_n + 1/2 \dt$.
Normally such evaluations are obtained by computing the right-hand side $\mc{Z}(z)$ of an equation of the form 
\begin{equation} \label{eqn:standard_form}
\dd{z}{t} = \mc{Z}(z) \, .
\end{equation}
However, as noted previously, the right-hand side from the explicit form (\cref{eqn:zeqn_exp}) is not accessible.
Instead, consider the following fixed-point problems for $dz / dt$ which are in the form of \cref{eqn:linear_fixedpoint}:
\begin{gather} 
\dd{z}{t}^n = \mc{L}(z^n)\dd{z}{t}^n + \mc{R}_z^n \label{eqn:dz_fixedpoint_stageone}\\
\dd{z}{t}' = \mc{L}(z')\dd{z}{t}' + \mc{R}_z' \label{eqn:dz_fixedpoint_stagetwo}\, .
\end{gather}
Assuming $\rho(\mc{L}) < 1$, \cref{eqn:dz_fixedpoint_stageone,eqn:dz_fixedpoint_stagetwo} are solved incompletely for $dz/dt$ with $k^f$ iterations of \cref{eqn:y_k} as discussed in \cref{subsec:incomp}.
The objectives of this calculation are
\begin{enumerate}
\item to obtain $\left(dz / dt\right)^{n,k^f}$ and $\left(dz / dt\right)^{',k^f}$ that are sufficiently accurate so $z$ has second-order convergence in time when time advanced with RK2.
The evaluation of the time derivative in RK2 is never more accurate than second order since it is computed from a second-order approximation of the solution, therefore $\dd{z}{t}$ need only be approximated to this accuracy.
\item to obtain $\left(dz / dt\right)^{n,k^f}$ and $\left(dz / dt\right)^{',k^f}$ that are sufficiently converged to mitigate instability (\cref{subsec:L}).
\end{enumerate}

To the first point, it is useful to employ a time history of $z$ to generate initial guesses for the fixed-point problems which can be made accurate to $\mc{O}\left(\dt^2\right)$. 
\todo{Mention somewhere (conclusion maybe) that this comes at the cost of more parasitic modes and third-order was not possible?}
As shown in \cref{eqn:y_k_orderep}, $\left(dz / dt\right)^{n,k^f}$ and $\left(dz / dt\right)^{',k^f}$ will then also be accurate to $\mc{O}\left(\dt^2\right)$.
To the second point, the number of iterations $k^f$ can be selected, for example, by considering the evolution of the absolute stability region for an appropriate test problem as a function of $k^f$ (see \cref{subsec:linear_stab}).

Backwards differentiation formula- (BDF)-like approximations using a time history including the current stage and the previous three stages provide the initial guesses for the fixed-point iterations that are accurate up to $\mc{O}\left(\dt^2\right)$.
They are
\begin{equation} \label{eqn:dzdt_guesses}
\begin{aligned}
\dt \dd{z}{t}^{n,0} &= 2 z^n - \left(2 + \dfrac{1}{2}\beta^n\right) z^{n-1,'} + \beta^n z^{n-1} - \dfrac{1}{2} \beta^n z^{n-2,'} \\
\dt \dd{z}{t}^{',0} &= \left(4 + \beta^{\prime}\right) z^{\prime} - \left(5 + 2 \beta^{\prime}\right) z^n + \beta^{\prime} z^{n-1,'} + z^{n-1} \, 
\end{aligned}
\end{equation}
where $\beta^n$ and $\beta^{\prime}$ are free parameters (see \cref{app:coeffs}).
Going forward we take $\beta^n = 0$ to limit the number of potential parasitic modes introduced by these approximations.

\subsection{A note on the curl-free momentum}
The solutions of \cref{eqn:dz_fixedpoint_stageone,eqn:dz_fixedpoint_stageone} $\left(dz / dt\right)^{n,k^f}$ and $\left(dz / dt\right)^{',k^f}$ are used to both time advance $z$ with RK2 and in the mass conservation equation to enforce the divergence constraint on $\bm{m}^c$:
\begin{equation}
\dd{f}{z}^n \dd{z}{t}^{n,k^f} = -\div{\bm{m}^{c,n}} \qquad \dd{f}{z}' \dd{z}{t}^{',k^f} = -\div{\bm{m}^{c,'}} \, .
\end{equation}
The solution of these equations is discussed in \cref{subsec:CF_recon} but we note here that the $\mc{O}\left(\dt^2\right)$ errors in $\left(dz / dt\right)^{n,k^f}$ and $\left(dz / dt\right)^{',k^f}$ lead to $\mc{O}\left(\dt^2\right)$ errors in $\bm{m}^c$.
Since $\bm{m}^c$ is obtained directly from the constraint rather than time advanced, the local, not global, error is relevant for accuracy so it will be temporally second-order accurate.
Furthermore, second-order errors in $\bm{m}^c$ will only contribute second-order errors to right-hand side evaluations. 
Hence, the errors in $\bm{m}^c$ contribute $\mc{O}\left(\dt^3\right)$ and higher errors to the time-advanced variables and the overall state is second-order accurate as desired.

\subsection{Explicit RK2 method} \label{subsec:time_scheme}
To introduce the full scheme and analyze its order of accuracy and stability, consider the following system of ODEs along with the equation of state, designed to exhibit the challenges arising in the low-Mach-number, variable-density equations:
\begin{equation} \label[system]{eqn:ODE}
\left\{\begin{aligned}
\dfrac{dz}{dt} = \mc{M}_1(z)&\dfrac{dz}{dt} + \mc{M}_2(z) \\
\dfrac{dm^d}{dt} = \mc{M}_3&(m^d,m^c) \\
\dfrac{d \rho}{dt}  = \mc{G}&(m^c) \\
\rho = f&(z)
\end{aligned} \right.
\end{equation}
where $\mc{M}_3(m^d,m^c)$ is potentially nonlinear in both $m^d$ and $m^c$ and is a surrogate for the momentum equations (\cref{eqn:COMom}).
$\mc{G}$ is a linear operator acting on $m^c$ akin to the divergence operator in the mass conservation equation.
In the $z$ equation, $\mc{M}_1$ is a potentially $z$-dependent linear operator acting on $dz/dt$ and $\mc{M}_2$ is generally nonlinear in $z$ -- this equation has the same structure as \cref{eqn:zeqn_imp} with $\mc{M}_1$ playing the role of $\mc{L}$ and $\mc{M}_2$ the role of $\mc{R}_z$.

For an arbitrary stage $s \in \left\{n,'\right\}$ and a chosen number of fixed-point iterations $k^f$, $(z^s,m^{d,s},m^{c,s})$ are updated by \cref{alg:time_advanceODE}. 
\begin{algorithm}[ht]\setstretch{2.}
\caption{Time advancement of \cref{eqn:ODE}}
\label{alg:time_advanceODE}
\begin{algorithmic}[1]
\item Start with $m^{d,s}, m^{c,s}, \dd{z}{t}^{s,k^f}$, $\dd{z}{t}^{s-1,k^f}$, $\mc{M}_3(m^{d,s-1},m^{c,s-1})$ from previous stages, evaluate $\mc{M}_3(m^{d,s},m^{c,s})$.
\item Advance $z^s \rightarrow z^{s+1}$, $m^{d,s} \rightarrow m^{d,s+1}$ with RK2.
\item Generate initial guess $\dd{z}{t}^{s+1,0}$ as shown in \cref{eqn:dzdt_guesses}.
\item $k = 0$
\For{$k <= k_f$}
	\State $\dd{z}{t}^{s+1,k} = \mc{M}_1(z^{s+1}) \dd{z}{t}^{s+1,k-1} + \mc{M}_2(z^{s+1})$
\EndFor
\item Evaluate $\dd{\rho}{t}^{s+1} = \dd{f}{z}(z^{s+1})\dd{z}{t}^{s+1,k_f}$, solve for $m^{c,s+1} = \mc{G}^{-1}\dd{\rho}{t}^{s+1}$.
\end{algorithmic}
\end{algorithm}
\subsubsection{Temporal convergence tests}
To verify the order of accuracy of the temporal scheme, numerical solutions of \cref{eqn:ODE} are sought with
\begin{equation}
\begin{aligned}
\mc{M}_1(z) = c_1 \left( 1 + \sin(z) \right) \, , \quad \mc{M}_2(z) &= c_2 \exp(z) \, , \quad \mc{M}_3(m^d,m^c) = c_3 m^d \left(m^{c}\right)^{3} \, , \\
\quad \mc{G}(m^c) = c_4 m^c \, , & \quad f(z) = z \, ,
\end{aligned}
\end{equation}
where $c_1 = .6\im$, $c_2 = -1$, $c_3 = 2 - .8\im$, $c_4 = -3 + .5\im$, and $z_0 = m^d_0 = .5$.
A reference solution is obtained by inverting $\mc{I} - \mc{M}_1$ directly and integrating
\begin{equation}
\dd{z}{t} = \dfrac{\mc{M}_2(z)}{1 - \mc{M}_1(z)} \, .
\end{equation}
To control temporal errors in the reference solution a fine $\Delta t = 1 \times 10^{-9}$ is used, an order of magnitude lower than the final test case shown here.
To ensure consistency of the initial condition for the test cases and reference solution, $m^c_0$ is specified as
\begin{equation}
m^c_0 = \mc{G}^{-1} \left[ \dd{f}{z} \dfrac{\mc{M}_2}{1 - \mc{M}_1} \right]_{z = z_0} \, .
\end{equation} 
The BDF parameters are set to $\beta' = -4$ and $\beta^n = 0$
(\cref{eqn:dzdt_guesses}) for generating the initial guesses for the fixed-point problems.

\begin{figure}[t]
  \centering
  \includegraphics[]{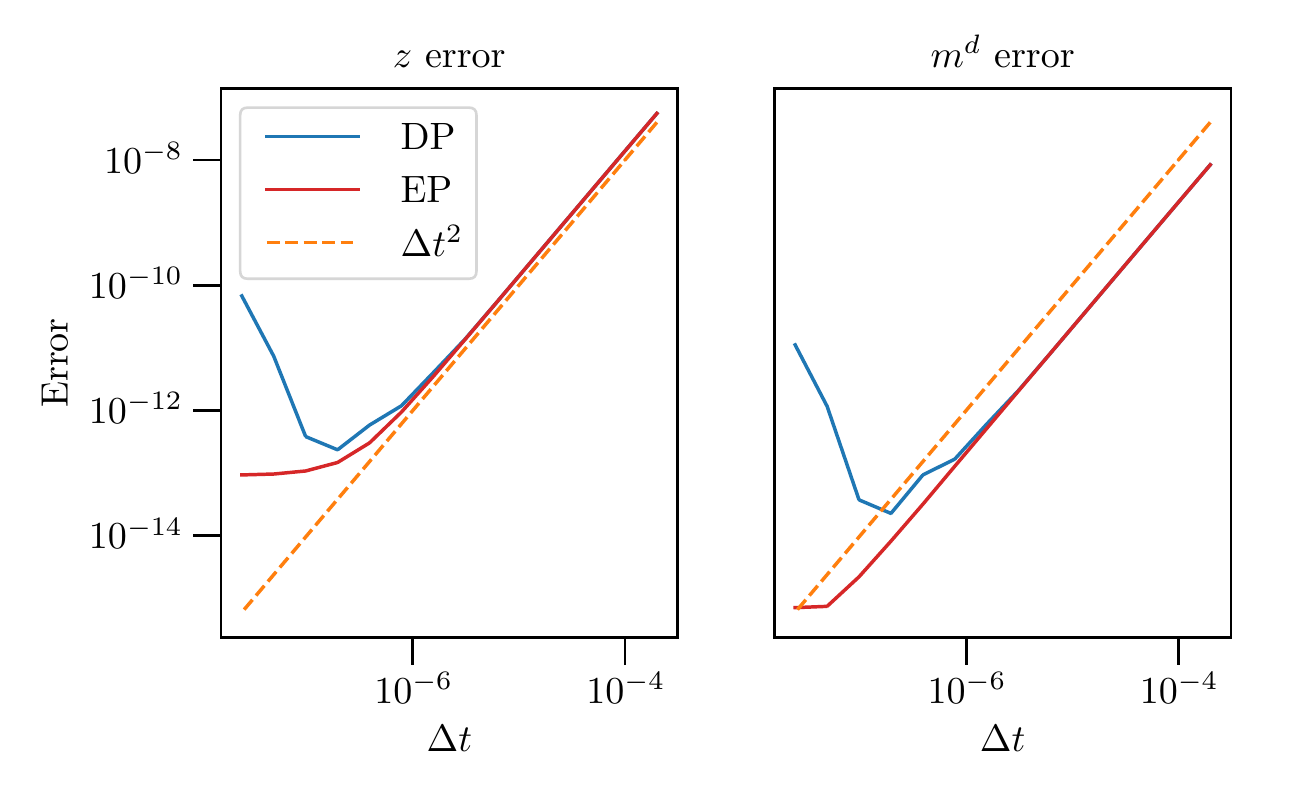}
  \caption[Temporal convergence of the iterative scheme]{Temporal convergence of the iterative scheme. $z$ (left), $m^d$ (right). Solution error at $t = .1$ defined against the reference case. Shown for both double precision (DP) and extended precision (EP).} \label{fig:convergence}
\end{figure}

\Cref{fig:convergence} shows the convergence rate of $z$ and $m^d$ for the test problem with both double and extended (long double) precision by comparing the solution at $t = .1$ to the reference solution.
Four iterations ($k^f = 4$) are used to solve the $z$ equation at each substep.
Some noise in the error is expected close to machine precision; however, the behavior seen for small $\Delta t$ needs further explanation.
Round-off errors in $z$ are amplified during the estimation of $dz/dt$ from the time history of $z$ because a linear combination of the four previous evaluations is scaled by $1/\dt$ (see \cref{eqn:dzdt_guesses}).
These errors affect $m^c$ and, in turn, can degrade the apparent convergence rate of $m^d$ and $z$.
In extended precision calculations second-order convergence continues to smaller $\dt$ (\cref{fig:convergence}), which demonstrates that the convergence floor in the double precision result is solely a consequence of round-off error.
The approximation of $m^c$ is thus confirmed to be consistent with the standard RK2 scheme and the convergence of the entire state is second order.
\subsubsection{Linear stability analysis} \label{subsec:linear_stab}
To assess the stability of this algorithm, consider a linearized system modeling the characteristics of the $z$ equation in \cref{eqn:ODE}
\begin{equation} \label{eqn:testODE_linear}
\dfrac{dz}{dt} = \lambda_1 \dfrac{dz}{dt} + \lambda_2 z
\end{equation}
where $\lambda_1$,$\lambda_2 \in \mathbb{C}$.
Additionally, define $\lambda^\ast \equiv \dfrac{\lambda_2}{1 - \lambda_1}$ so that
\begin{equation} \label{eqn:testODE_linear_collapsed}
\dfrac{dz}{dt} = \lambda^\ast z
\end{equation}
is the underlying modified ODE (assuming $\lambda_1 \neq 1$).
In this context, setting $\lambda_2 = 0$ in \cref{eqn:testODE_linear} gives a test problem for zero stability.

Applying the temporal scheme with $k$ fixed-point iterations to \cref{eqn:testODE_linear} gives
\begin{equation}
\begin{aligned}
{\zso}^{,k} - z^n &= \lambda_1^k \left\{ z^n - z^{n-1,'} \right\} + \dfrac{1}{2} \dt \sum_{j=0}^{k-1} \lambda_1^j  \lambda_2 z^n \\
{\zno}^{,k} - \zso &= \lambda_1^k \left\{ \left(3 + \beta'\right) \zso - \left(6 + 2\beta'\right) z^n + \left(\beta' + 1\right) z^{n-1,'} + z^{n-1} \right\} \\
&\quad\quad + \dt \sum_{j=0}^{k-1} \lambda_1^j  \lambda_2 \left( \zso - \dfrac{1}{2} z^n \right) \, .
\end{aligned}
\end{equation}
%
Taking inspiration from linear differential equation analysis \cite{Mallik1998}, this can be written as $\bm{z}^{n+1} = \bm{A}_1 \bm{A}_2 \bm{z}^n = \bm{A} \bm{z}^n$
%
%
where $\bm{z}^{n} = [z^n \quad z^{n-1,'} \quad z^{n-1} \quad z^{n-2,'}]^T$ (the matrices are detailed in \cref{app:lstab}).
%
%
For $\rho\left(\bm{A}\right) < 1$, the scheme is linearly stable.
Unlike traditional explicit Runge-Kutta methods applied to ODEs of the form $dz/dt = f(z)$ (like \cref{eqn:testODE_linear_collapsed}), the scheme is not zero stable.
The zero stability is instead like that of a linear multistep method due to the parasitic modes introduced by the data required to obtain the initial guesses for the fixed-point scheme.
In this case, $\lambda_1$ has a direct impact on zero stability.
\begin{figure}[t]
  \centering
  \includegraphics[]{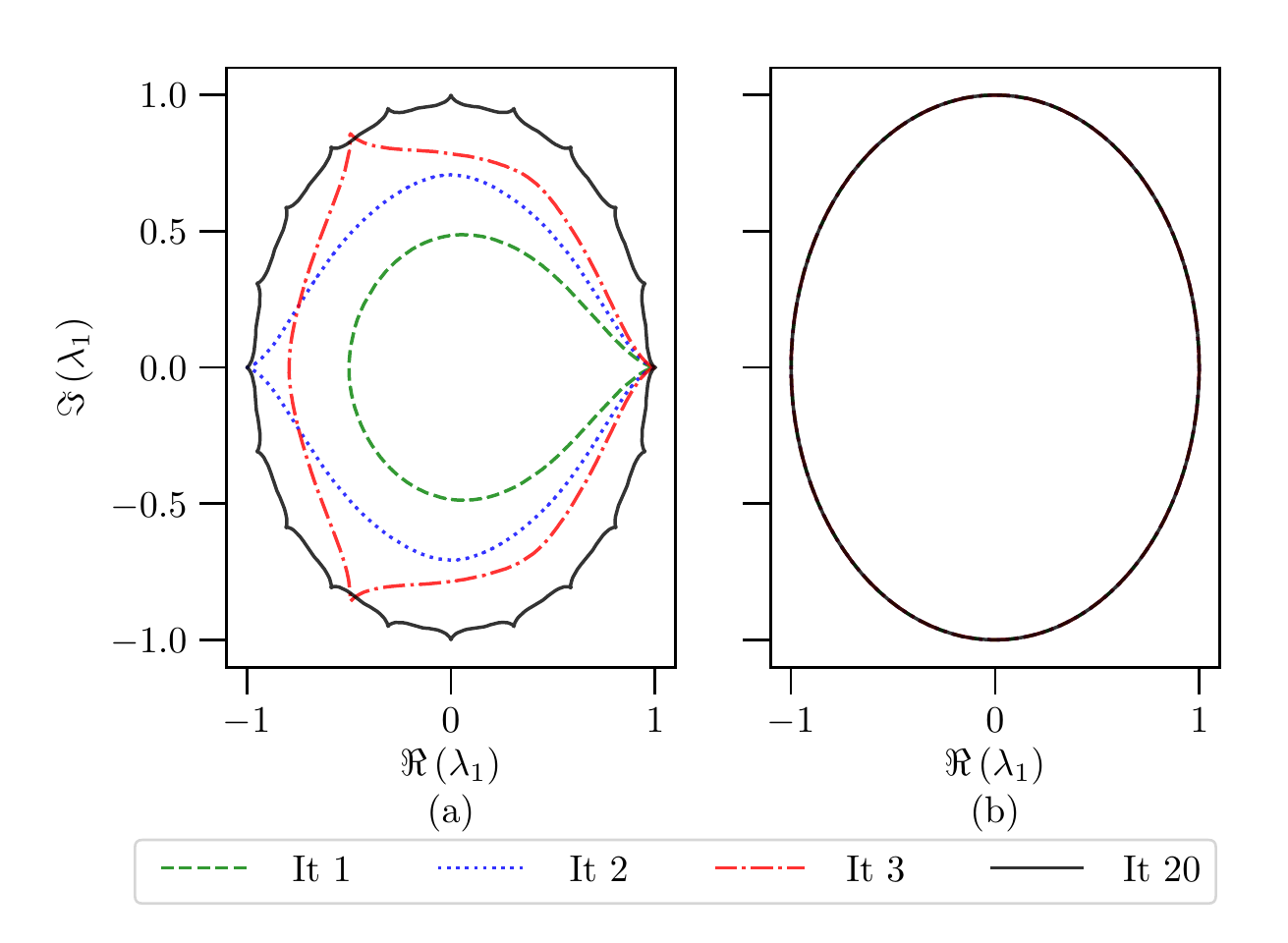}
  \caption[Zero stability of the iterative scheme]{Zero stability of the iterative scheme for the first few iterations. \textit{Left}: $\beta' = -1$, iteration 20 shows a nearly converged region. \textit{Right}: $\beta' = -4$, the optimal region is obtained for all iterations.} \label{fig:zero_stab}
\end{figure}
Generally, the size of the zero stability regions grows with the number of iterations for a given $\beta'$ (\cref{fig:zero_stab}.(a)).
When fully converged, the zero-stability region is a circle of radius $1$, which is consistent with the scheme diverging for $\rho\left(\mc{M}_1\right) \geq 1$ and recovering the stability of the RK2 scheme otherwise.
Setting $\beta' = -4$ yields the converged zero-stability region regardless of the number of iterations (\cref{fig:zero_stab}.(b)), therefore this choice has been used to obtain all other results herein.

\begin{figure}[t]
  \centering
  \includegraphics[]{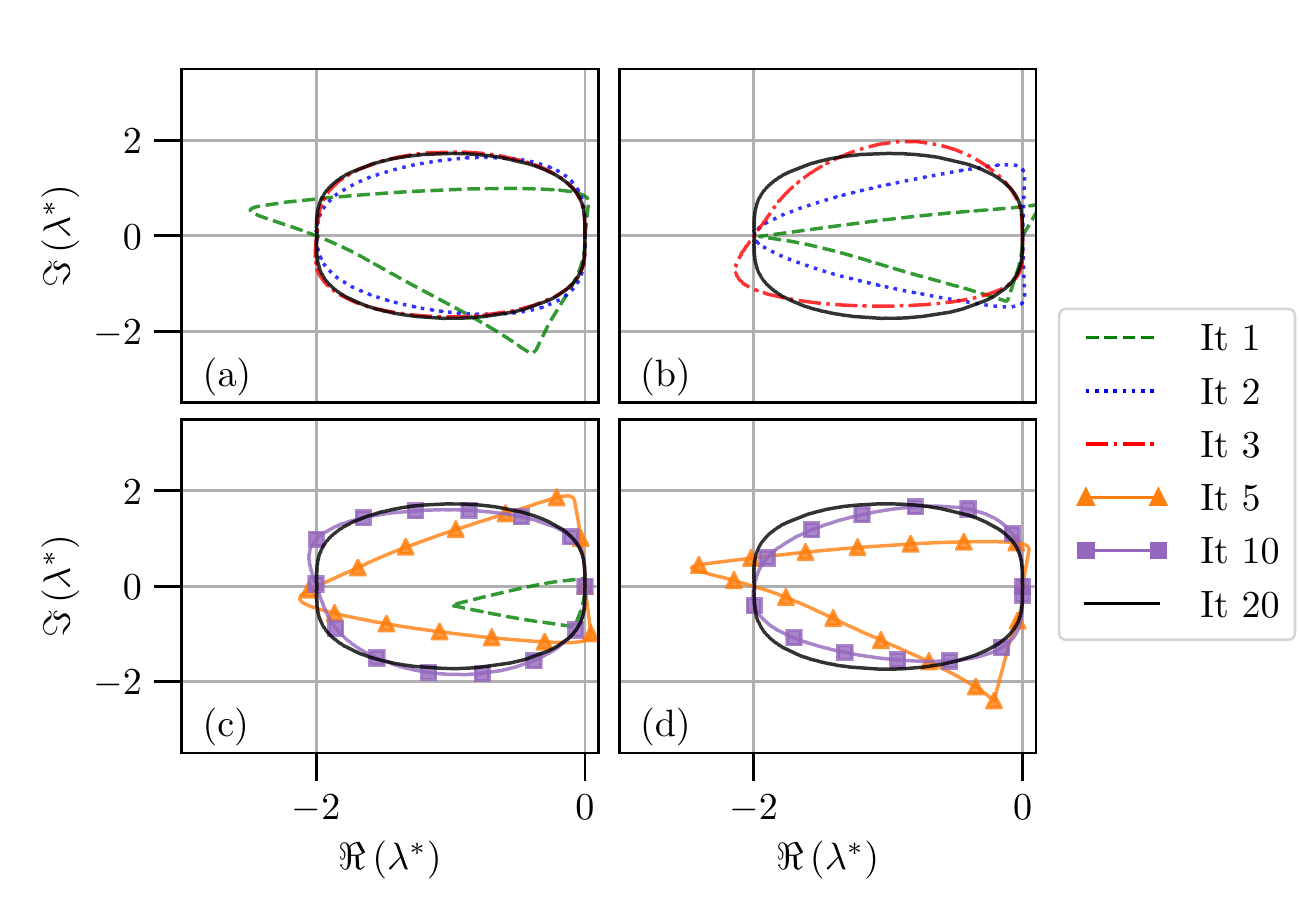}
  \caption[Absolute stability as a function of the number of iterations for varying $\lambda_1$]{Absolute stability as a function of the number of iterations for varying $\lambda_1$. \textbf{(a)}: $\lambda_1 = .2\im$ (iters.~5, 10 not shown), \textbf{(b)}: $\lambda_1 = .4\im$ (iters.~5, 10 not shown), \textbf{(c)}: $\lambda_1 = .5 + .5\im$ (iters. 2,3 not shown), \textbf{(d)}: $\lambda_1 = .7 + .25\im$ (iters.~2,3 not shown -- not stable for 1 iter.).} \label{fig:abs_stab_grow}
\end{figure}

The absolute-stability region is defined in terms of the ``true'' eigenvalue, $\lambda^\ast$, and is plotted for different values of $\lambda_1$, see \cref{fig:abs_stab_grow}.
\begin{figure}[t]
  \centering
  \includegraphics[]{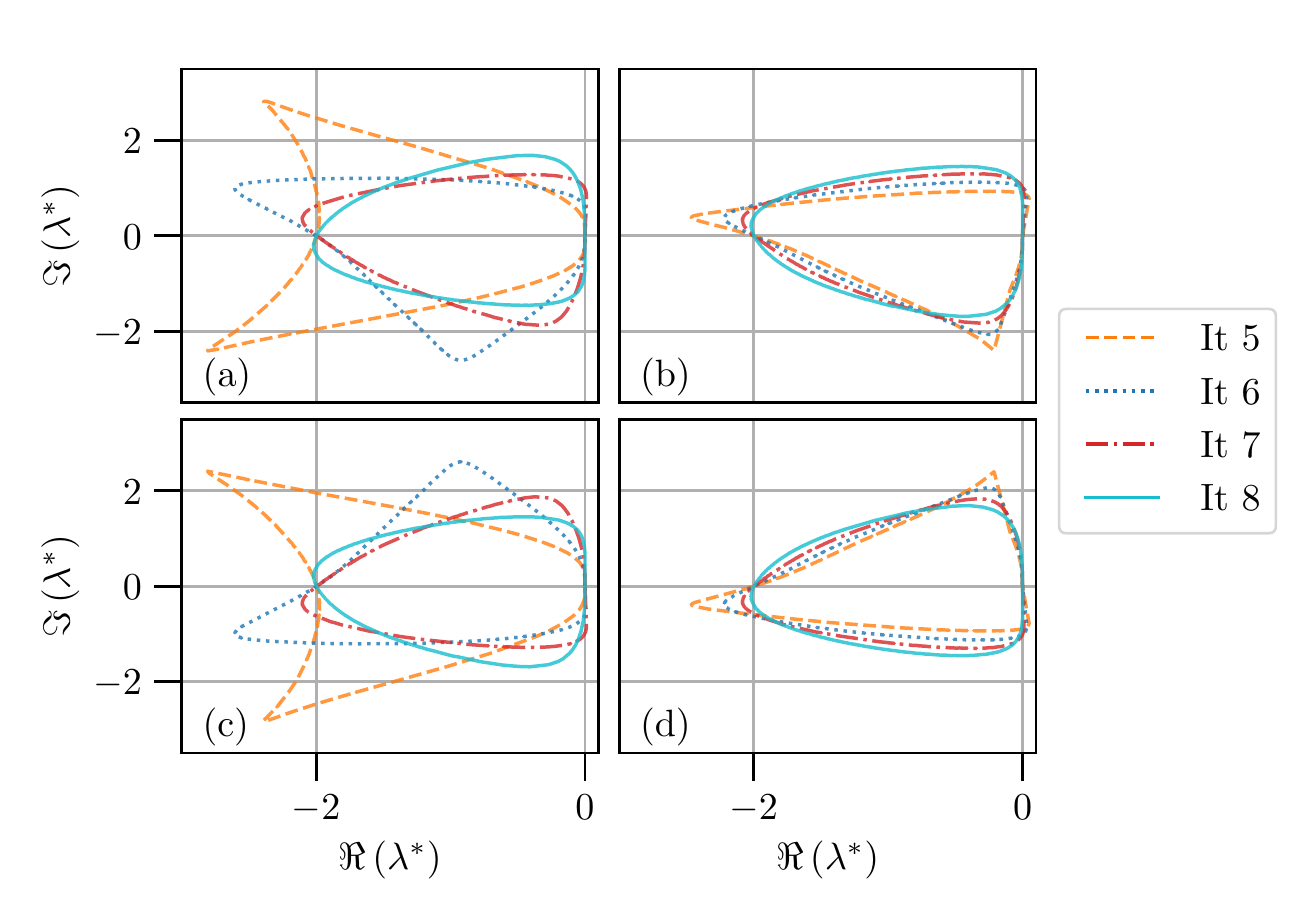}
  \caption[Absolute stability as a function of the number of iterations for $\lambda_1$ of equal magnitude but different phase]{Absolute stability as a function of the number of iterations of the for $\lambda_1$ of equal magnitude but different phase. \textbf{(a)}: $\lambda_1 = .25 + .7 \im$, \textbf{(b)}: $\lambda_1 = .7 + .25\im$, \textbf{(c)}: $\lambda_1 = .25 - .7\im$, \textbf{(d)}: $\lambda_1 = .7 - .25\im$.} \label{fig:abs_stab_sym}
\end{figure}
As expected, when fully converged the absolute stability region coincides with the standard RK2 result.
For small $\| \lambda_1 \|$ this convergence happens quickly so that only two or three iterations are required for a stability region that approaches that of RK2.
As the eigenvalue magnitude approaches one, as in \cref{fig:abs_stab_grow}.d, around ten iterations are needed for the stability region to near that of RK2.
\Cref{fig:abs_stab_sym} shows the effect of the phase of $\lambda_1$ on the stability region.
In particular, there is a symmetry about the imaginary axis when $\mathbb{I}\left[\lambda_1\right]$ changes sign.
Additionally, a $180^\circ$ phase change can produce dramatic changes in stability as the eigenvalues with large imaginary parts tend to exhibit more erratic behavior.
The type of analysis shown here can be used as a heuristic to determine an appropriate number of iterations for the fixed-point problem if an estimate of the largest eigenvalues of $\mc{L}$ are obtained $\textit{a priori}$.
\FloatBarrier

%% file: algo.tex
%
In this section we demonstrate the utility of decomposing the momentum (as in \cref{eqn:mom_decomp}) and summarize the advancement of the state over one timestep.
The DNS numerical technique described here is appropriate for simulation domains with periodic boundary conditions in two directions.
The computational domain is $\mc{V} = [0,L_1] \times [-L_2/2,L_2/2] \times [0,L_3]$.
A general flow variable $f$ is expressed as a Fourier series in $x_1$ and $x_3$:
\begin{equation} \label{eqn:disc_f}
\begin{aligned}
f(x_1,x_2,x_3,t) &=  \sum_{l = - \frac{N_1}{2} + 1}^{\frac{N_1}{2} - 1}
				 \sum_{n = - \frac{N_3}{2} + 1}^{\frac{N_3}{2} - 1}
				 \widehat{f}_{ln}(x_2,t) e^{\im k_1 x_1}
				 e^{\im k_3 x_3} \\
				 &\quad k_1 = \dfrac{2\pi l}{L_1} \quad\quad k_3 = \dfrac{2\pi n}{L_3}
\end{aligned}
\end{equation}
Note that the dependence of $k_1$ and $k_3$ on $l$ and $n$ is assumed, though not explicitly indicated, and spatial discretization in the inhomogeneous direction is deferred.
However, we note that the algorithm was designed to guarantee discrete conservation of mass when an appropriate basis for the $x_2$ direction is chosen (discussed below and in \cref{app:spat_disc}).
\subsection{The divergence-free momentum} \label{subsec:md_recon}
Inspired by the formulation of Kim, Moin, and Moser \cite{kim1987turbulence} (for incompressible flow), define
\begin{equation}
\begin{aligned}
\bm{\phi} \equiv \lap{\bm{m}^d} \quad \quad \bm{\Omega} \equiv \nabla \times \bm{m}^d \, .
\end{aligned}
\end{equation}
The $2$-component of the curl and double curl of the momentum equations \eqref{eqn:COMom} are then given by
%
\begin{gather}
\pp{\Omega_2}{t} = \pspp{}{x_3}{x_j} \left ( C_{1j} + \tau_{1j} \right ) - \pspp{}{x_1}{x_j} \left ( C_{3j} + \tau_{3j} \right ) \label{eqn:om} \\
\pp{\phi_2}{t} = \dfrac{\partial^3}{\partial x_l \partial x_l \partial x_j} \left (C_{2j} +\tau_{2j} \right ) - \dfrac{\partial^3}{\partial x_2 \partial x_l \partial x_j} \left ( C_{lj} + \tau_{lj} \right ) \, , \label{eqn:phi} \\ 
\Omega_2 = \pp{m_1^d}{x_3} - \pp{m_3^d}{x_1} \quad\quad \phi_2 = \lap{m_2^d} \notag
\end{gather}
%
with $C_{ij} \equiv - \rho u_i u_j$.
The $2$-component is used here because the $x_1$ and $x_3$ directions are treated with Fourier spectral methods as in KMM.
Since \cref{eqn:om,eqn:phi} were obtained by applying the curl operators to the momentum equations, they leave $\bm{m}^d$ averaged in $x_1$ and $x_3$ undetermined.
Averaging the momentum equations over these directions and denoting such planar averages as $\pa{\hspace{4pt}\cdot\hspace{4pt}}$ gives:
\begin{equation} \label{eqn:mean_m1m3}
\begin{aligned}
\pp{\hspace{1pt}\pa{m}_1}{t} &= \pp{}{x_2}\left(\pa{C}_{12} + \pa{\tau}_{12}\right) \\
\pp{\hspace{1pt}\pa{m}_3}{t} &= \pp{}{x_2}\left(\pa{C}_{23} + \pa{\tau}_{23}\right) 
\end{aligned}
\end{equation}
where $\pa{m}_1 = \pa{m^d}_1$, $\pa{m}_3 = \pa{m^d}_3$, and $\pa{m^d}_2 = 0$.
The fact that $\partial \pa{\tau}_{ij} / \partial x_j$ simplifies to $\partial \pa{\tau}_{i2} / \partial x_2$ has also been used.

\Cref{eqn:om,eqn:phi,eqn:mean_m1m3} govern the evolution of the divergence-free momentum.
The main advantages of recasting the momentum equations in this way is that the pressure is eliminated and the divergence-free condition can be imposed exactly.
For the modes with one or more non-zero wavenumbers, $m_2^d$ is obtained in Fourier space from the Poisson problem
\begin{equation} \label{eqn:phi_poisson}
\hpvC = -k^2 \hmdyC{2} + \psps{\hmdyC{2}}{x_2}
\end{equation}
where $k^2 = k_1^2 + k_3^2$ and we have suppressed the $l$ and $n$ indices for compactness.
Then the $x_1$ and $x_3$ components are reconstructed by invoking the divergence-free condition and the definition of $\Omega_2$:
\begin{equation} \label{eqn:md_recon}
\begin{aligned}
\hmdyC{1} &= \dfrac{1}{k^2}\left(\im k_1 \pp{\hmdyC{2}}{x_2} - \im k_3 \hovC \right) \\
\hmdyC{3} &= \dfrac{1}{k^2}\left(\im k_3 \pp{\hmdyC{2}}{x_2} + \im k_1 \hovC \right) \, .
\end{aligned}
\end{equation}
Provided the numerical representation of the $x_2$ dependence of $\hmdyC{1}$, $\hmdyC{3}$ and $\hovC$ can exactly represent the $x_2$ derivative of $\hmdyC{2}$, this procedure will produce $\bm{m}^d$ that is exactly divergence free.
\subsection{The curl-free momentum and discrete conservation of mass} \label{subsec:CF_recon}
Determining $\bm{m}^c$ from the mass conservation equation appears straightforward. 
We need only find a vector field $\bm{m}^c$ that satisfies two constraints:
\begin{align}
  \div{\bm{m}^c} &=-\pp{\rho}{t} \label{eqn:constraint1} \\
  \nabla\times\bm{m}^c &=0 \, . \label{eqn:constraint2}
\end{align}
Analytically, this is easily accomplished by writing $\bm{m}^c=\nabla\psi$, which ensures that \cref{eqn:constraint2} is satisfied, and then solving a Poisson equation for $\psi$ derived from \cref{eqn:constraint1} (i.e.~\cref{eqn:COM_with_psi}). 
However, when using this approach numerically, there will generally be error in the solution of the Poisson equation -- that is the divergence of the gradient of $\psi$ will not be exactly $-\partial\rho/\partial t$, introducing mass conservation errors. 
At the same time, depending on the details of the numerics, the discrete curl of the discrete gradient will often be exactly zero, and indeed that would be the case for the discretizations used here (\cref{app:spat_disc}). 
Errors in mass conservation can cause significant problems, and so we seek a solution strategy for $\bm{m}^c$ that can exactly satisfy \cref{eqn:constraint1}, perhaps at the expense of errors in \cref{eqn:constraint2}.

The mass conservation equation only involves the curl-free momentum in
the $x_2$ direction ($m_2^c$) and in the direction of the wavevector
($k_1 m_1^c + k_3 m_3^c = \bm{k} \cdot \bm{m}^c$). Mass conservation
and the curl-free condition can thus be written in Fourier space as:
\begin{gather}
\widehat{\dd{f}{z}\pp{z}{t}} = \pp{\hryC}{t} = -\im
k \widehat{m^c_{\parallel}} - \pp{\hmcyC{2}}{x_2}  \qquad\mbox{mass conservation}\label{eqn:COM_fourier} \\
\pp{\widehat{m^c_{\parallel}}}{x_2} - \im k \widehat{m^c_2} = 0 \qquad\mbox{curl-free}\label{eqn:CF_fourier}
\end{gather}
where $m^c_\parallel \equiv \bm{k} \cdot \bm{m}/k$ .
%
A second curl-free constraint, $k_3 \hmcyC{1} - k_1 \hmcyC{3} = 0$,
along with the definition of $\widehat{m^c_\parallel}$ allows $\hmcyC{1}$
and $\hmcyC{3}$ to be determined from $\widehat{m^c_\parallel}$ as
\begin{equation} \label{eqn:mc_recon}
\begin{aligned}
\hmcyC{1} = &\dfrac{k_1}{k}\widehat{m^c_\parallel} \quad \quad \hmcyC{3} = \dfrac{k_3}{k} \widehat{m^c_\parallel} \, .
\end{aligned}
\end{equation}
Note that \cref{eqn:COM_fourier,eqn:CF_fourier} can be combined to get a second order ODE in either $\widehat{m^c_\parallel}$ or $\widehat{m^c_2}$. 
But if they were to be solved this way we would lose independent control over the discretization error in each equation. 
To ensure that mass conservation is satisfied exactly, we use a representation for $\hmcyC{2}$ whose $x_2$ derivative is exactly represented in the function space used to represent $\widehat{m^c_\parallel}$, $\widehat{\rho}$ etc., and then require that (\cref{eqn:COM_fourier}) be satisfied without discretization error. 
The coupled equations~(\ref{eqn:COM_fourier}-\ref{eqn:CF_fourier}) are thus solved together along with the required boundary conditions on $\hmcyC{2}$ (see \cref{subsec:bcs}), with discretization error only in (\ref{eqn:CF_fourier}). 
Also analogous to \cref{subsec:md_recon}, the planar averages of $\bm{m}^c$ must be determined as a special case, that is
\begin{equation} \label{eqn:mc_mean}
\pp{\pa{m^c}_2}{x_2} = -\pp{\pa{\rho}}{t} \qquad \pa{m^c}_1 = \pa{m^c}_3 = 0 \, .
\end{equation}
\subsection{Applying $\mc{L}$ for the fixed-point problem}
Let $\mc{C}: \partial \rho / \partial t \rightarrow \bm{m}^c$ be the operator that maps $\partial \rho / \partial t$ to $\bm{m}^c$ as described in \cref{subsec:CF_recon} (see \cref{eqn:constraint1,eqn:constraint2}), then $\mc{L}(z)$ can be written
\begin{equation}
\mc{L}(z)\left\{\cdot\right\} = - \dfrac{1}{\rho} \grad{z} \cdot \mc{C} \left[\dd{f}{z}\{\cdot\}\right] \, .
\end{equation}
Practically, $\mc{L}$ is applied to $\partial z / \partial t$ by first forming the nonlinear product
\begin{equation}
\pp{\rho}{t} = \dd{f}{z}\pp{z}{t} \, , 
\end{equation}
then solving for the curl-free momentum as discussed in \cref{subsec:CF_recon}, and finally computing the convective term
\begin{equation}
- \dfrac{1}{\rho} \grad{z} \cdot \bm{m}^c \, .
\end{equation}
\subsection{Time advance}
This section details the time advance of the evolution equations with the temporal scheme established in \cref{sec:temp_disc}.
The final DNS equations are
\begin{equation} \label[system]{eqn:DNS_system}
\begin{gathered}
-\dd{f}{z}\pp{z}{t} = \pp{m^c_j}{x_j} \\
\pp{\Omega_2}{t} = \pspp{}{x_3}{x_j} \left ( C_{1j} + \tau_{1j} \right ) - \pspp{}{x_1}{x_j} \left ( C_{3j} + \tau_{3j} \right ) = \mc{RHS}_{\Omega_2} \\
\pp{\phi_2}{t} = \dfrac{\partial^3}{\partial x_l \partial x_l \partial x_j} \left (C_{2j} + \tau_{2j} \right ) - \dfrac{\partial^3}{\partial x_2 \partial x_l \partial x_j} \left ( C_{lj} + \tau_{lj} \right ) = \mc{RHS}_{\phi_2} \\
\pp{\hspace{1pt}\pa{m}_1}{t} = \pp{}{x_2}\left(\pa{C}_{12} + \pa{\tau}_{12}\right) = \mc{RHS}_{\pa{m}_1} \\
\pp{\hspace{1pt}\pa{m}_3}{t} = \pp{}{x_2}\left(\pa{C}_{23} + \pa{\tau}_{23}\right) = \mc{RHS}_{\pa{m}_3} \\
\pp{z}{t} = \mc{L}(z)\pp{z}{t} + \mc{R}_z(z) \\
\rho = f(z) \, .
\end{gathered}
\end{equation}

Throughout the calculation, operations are performed on the data in either Fourier space (also referred to as `wavespace'), when the coefficients are stored and the data is ordered by wavenumber pair, or in `realspace', when it is evaluated on a physical grid via discrete Fourier transforms.
Time advancement and linear algebra (including differentiation) is performed in wavespace on the coefficients, and nonlinear operators are computed in realspace to avoid costly convolution sums.
An earlier version of this algorithm was implemented for a combustion DNS using B-splines for the $x_2$ representation in \cite{reuter2021}, which contains details on optimizing for storage requirements and minimizing data movement.

The time advance of the state $\bm{\mc{S}} = \left(\phi_2,\Omega_2,z,\pa{m}_1,\pa{m}_3\right)$ over the course of one substep $s \rightarrow s+1$ is outlined in \cref{alg:time_advanceDNS}.
\Cref{alg:time_advanceODE} will be extended as appropriate to the spatially discrete problem defined by \cref{eqn:DNS_system}.
It is assumed 
\begin{equation*}
\mc{RHS}_{\Omega_2}^{s-1}, \; \mc{RHS}_{\phi_2}^{s-1}, \; \mc{RHS}_{\pa{m}_1}^{s-1}, \; \mc{RHS}_{\pa{m}_3}^{s-1}, \; \pp{z}{t}^{s-1,k_f}
\end{equation*}
are stored from the previous stage as well as the time history of $z$: ($z^s, z^{s-1}, z^{s-2}, z^{s-3}$) required for initializing the fixed-point problems.
\begin{algorithm}[ht]\setstretch{2.}
\caption{Time advancement of \cref{eqn:DNS_system}}
\label{alg:time_advanceDNS}
\begin{algorithmic}[1]
\item Finalize $\bm{m}^{d,s}$ by solving \cref{eqn:phi_poisson} and applying \cref{eqn:md_recon}.
\item Evaluate $\mc{R}_z^s$, generate BDF-like approximation of $\pp{z}{t}^{s,0}$.
\item Obtain $\pp{z}{t}^{s,k_f}$ as in \cref{subsec:fixedpoint_for_dns} then solve for $\bm{m}^c$ with \cref{eqn:COM_fourier,eqn:CF_fourier,eqn:mc_recon,eqn:mc_mean}.
\item Update thermodynamic and transport properties ($\rho^s$, $\mu^s$, $\mc{D}_z^s$) which are known in terms of $z^s$.
\item Evaluate $\mc{RHS}_{\Omega_2}^{s}, \; \mc{RHS}_{\phi_2}^{s}, \; \mc{RHS}_{\pa{m}_1}^{s}, \; \mc{RHS}_{\pa{m}_3}^{s}$.
\item Time advance the state $\mc{S}^s \rightarrow \mc{S}^{s+1}$ with RK2.
\end{algorithmic}
\end{algorithm}
\subsection{Boundary conditions} \label{subsec:bcs}
Unbounded domains which must be truncated for computational purposes
are commonly encountered in the simulation of turbulent flows, and
that is the case here for the inhomogeneous spatial direction, in which
the domain is formally infinite.  In
the homogeneous directions, periodic boundary conditions are employed
consistent with the use of Fourier expansions.  In the
inhomogeneous direction, the momentum boundary condition is an
extension of the potential-matching condition of Corral and
Jim\'{e}nez originally developed for incompressible
flows \cite{corral1995fourier}.  This is based on the assumption that
the vorticity decays rapidly as $x_2 \rightarrow \pm \infty$, so that
at the boundary of a truncated computational domain the potential part
of the velocity is consistent with a decaying irrotational solution in
the exterior.  In the variable-density case this means that, for $L_2$
sufficiently large, $\rho \rightarrow \rho_\infty$,
$\nabla \times \bm{m} \rightarrow 0$, and $\div{\bm{m}} \rightarrow
0$.

Consider a potential-only momentum field for $x_2 > L_2/2$, so that
$\bm{m} = \grad{\psi}$. Then $\psi$ obeys $\lap{\psi} = 0$ and the
Fourier coefficients of $\psi$ satisfy $\widehat{\psi}\sim e^{-kx_2}$,
except when $k=0$. Evaluating at the boundary of the
computational domain $x_2=L_2/2$, yields the potential-matching
boundary conditions for the top boundary, and a similar analysis
yields analogous conditions for the bottom boundary. They are:
\begin{equation}\label{eqn:bc_both}
\begin{aligned}
\left.\widehat{m_2}\right|_{\pm\frac{L_2}{2}} &= \mp\dfrac{1}{k}\pp{\hmpyC{2}}{x_2}\bigg|_{\pm\frac{L_2}{2}}\\
\left.\widehat{m_1}\right|_{\pm\frac{L_2}{2}} &= \mp\dfrac{\im k_1}{k}\left.\widehat{m_2}\right|_{\pm\frac{L_2}{2}}\\
\left.\widehat{m_3}\right|_{\pm\frac{L_2}{2}} &= \mp\dfrac{\im k_3}{k}\left.\widehat{m_2}\right|_{\pm\frac{L_2}{2}}
\end{aligned}
\end{equation} 
Clearly, both the divergence-free and curl-free components of $\bm{m}$
can satisfy these potential-matching condition individually, ensuring
that $\bm{m}$ does as well.

Recall that $\lap{m_2^d} = \phi_2$, so \cref{eqn:phi} is fourth order in $m_2^d$.
Two additional boundary conditions on $m_2^d$ are therefore needed.
From the potential-matching condition, it follows that
\begin{equation} \label{eqn:md_bc}
k^2\left.\widehat{m^d_2}\right|_{\pm\frac{L_2}{2}}- \left.\dfrac{\partial^2 \hmdyC{2}}{\partial x_2^2}\right|_{\pm\frac{L_2}{2}} = \left.\widehat{\phi_2}\right|_{\pm\frac{L_2}{2}}=0,\\
\end{equation}
A homogeneous Dirichlet boundary condition is thus applied to
$\phi_2$.  Then, when $m_2^d$ is reconstructed from $\phi_2$ via the
Poisson equation \cref{eqn:phi_poisson}, only the Robin conditions
of \cref{eqn:bc_both} are explicitly enforced.  A homogeneous
Dirichlet condition is enforced on $\Omega_2$ which is consistent with
curl-free momentum at the boundary.  When
solving \cref{eqn:COM_fourier,eqn:CF_fourier} for $m^c_2$ and
$m^c_\parallel$, the Robin conditions from \cref{eqn:bc_both} are
imposed on $\widehat{m^c_2}$.

Averaging the mass conservation equation over $x_1$ and $x_3$ yields a homogeneous Neumann condition on $\pa{m}_2$ at the boundary, since the flow is essentially constant density there.
This means there is one remaining degree of freedom when determining $\pa{m}_2$ which is used to enforce a symmetry condition or set the value at a location in the domain, for example. 
For the Rayleigh-Taylor problem, simulations are stopped well before the front nears the boundary so a homogeneous Dirichlet condition is used at the top of the domain as the fluid remains at rest.

Homogeneous Neumann conditions are imposed on the streamwise and spanwise plane-averaged momentum as well as the plane-averaged and fluctuating components of the transported scalar, $z$.
%

%% file: results.tex
This section contains numerical results using the
algorithm described here.  A single-mode Rayleigh-Taylor (RT)
instability is simulated over a range of Atwood numbers (density
ratios) to verify that the algorithm is obtaining solutions to the variable-density equations and that it is stable at
large density ratios.  Additionally, the implementation has been
verified (results not shown) with a manufactured solution created with
MASA \cite{malaya2013masa}, a C++ library that generates source terms
for arbitrary differential equations by automatic differentiation.

%
%
Following the setup in,
e.g., \cite{He1999,hamzehloo2021direct,lee2013numerical} of a
single-mode Rayleigh-Taylor instability in a rectangular box with
square cross section, we consider two fluids of differing
density arranged with the heavier fluid on top.  
They evolve under the influence of gravity, which is aligned with the vertical direction
($x_2$).  The dynamic viscosity of the two fluids is taken to be the
same as in \cite{hamzehloo2021direct} and to account for gravity, a body
force is added to the momentum equation in (\ref{eqn:LMNS_system}) as follows:
\begin{equation}
\pp{\rho u_i}{t} + \pp{\rho u_i u_j}{x_j} = - \pp{p}{x_i}
+ \pp{\tau_{ij}}{x_j} -\rho g\delta_{i2}
\end{equation}
We assume the fluids are miscible and simple binary, Fickian diffusion holds such that
\begin{equation}
z=\dfrac{\rho_h}{\rho_h-\rho_l}\left(1-\dfrac{\rho_l}{\rho}\right) \, ,
\end{equation}
which represents the volumetric mixing of the heavy and light fluids, satisfies the convection-diffusion equation in \cref{eqn:LMNS_system}. \
The equation of state, which relates the density to
$z$, is then
\begin{equation}
\rho=\dfrac{\rho_l\rho_h}{\rho_h-(\rho_h-\rho_l)z} \, .
\end{equation}
The Atwood number characterizes the density contrast between the heavy fluid, $\rho_h$, and light fluid, $\rho_l$, and is given by
\begin{equation}
  At = \dfrac{\rho_h - \rho_l}{\rho_h + \rho_l} \implies \dfrac{\rho_h}{\rho_l}=\dfrac{1+At}{1-At}.
\end{equation}

Computations are performed in both two and three dimensions, in a
horizontally periodic domain of size $W$ or $W\times W$. In the
vertical ($x_2$) direction the domain is $x_2\in[-2W,2W]$ or
$[-3.5W,3.5W]$, with the larger domains used for high Atwood number
($At$) cases.

A Fourier representation with $128$ modes is used in the periodic
directions ($x_1$ and $x_3$), and in the $x_2$ direction, either $512$
(smaller domain) or $1024$ (larger box) B-spline degrees of freedom
are employed (see \cref{app:mesh_conv} for a mesh convergence study).
Note that the resolution in the large and small domains is
approximately equal. In general, the timestep is chosen such that
$\dt \sqrt{At} / \sqrt{W/g} = 2.5 \times 10^{-4}$, although there are
a few exceptions for higher Atwood number cases due to viscous
stability constraints.
\Cref{app:mesh_conv} includes the grid and timestep for each case presented here.

\begin{figure}[!h]
  \centering
  \includegraphics[]{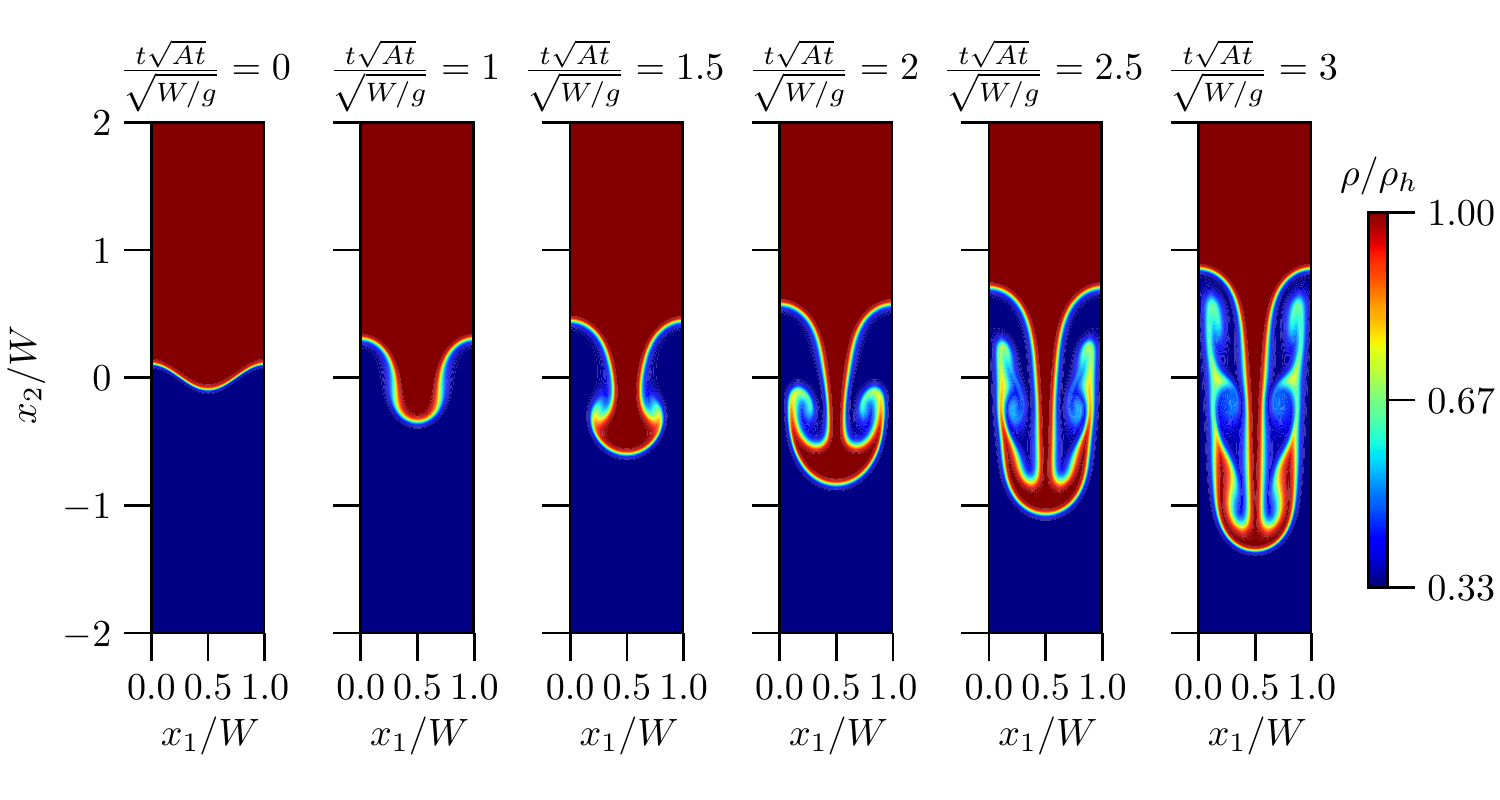}
  \caption[justification=raggedright]{Contours of density at several
  times for the two-dimensional RT problem with $At = .5$.} \label{fig:rt_2D_cont_At_.5}
\end{figure}

The initial $z$ is given by
\begin{equation}
z = \dfrac{1}{2} \left[ 1 + \tanh\left(\dfrac{x_2 - h}{2 \epsilon
W}\right) \right] +\delta
\end{equation}
where $\epsilon=0.05$ and $h$ is one of:
\begin{gather}
 \dfrac{h^{3D}(x_1,x_3)}{W} = 0.05 \left [ \cos\left(\dfrac{2\pi
 x_1}{W}\right) + \cos\left(\dfrac{2\pi x_3}{W}\right) \right] + \delta\\
 \dfrac{h^{2D}(x_1)}{W} = 0.1 \left [ \cos\left(\dfrac{2\pi
 x_1}{W}\right) \right] + \delta
\end{gather}
for the three- and two-dimensional cases, respectively.  At low $At$,
when the smaller vertical domain is used, the offset $\delta=0$, but
at higher $At$, when the larger domain is used, $\delta=0.5$. The high
$At$ vertical offset is used because the heavy fluid penetrates more
rapidly into the light fluid. An example of the initial $z$ field can be
seen in the leftmost panel of \cref{fig:rt_2D_cont_At_.5}.

In the cases computed here the Schmidt number is taken to be unity ($Sc = 1$), so in \cref{eqn:LMNS_system}, $\rho \mathcal{D}_z= \mu$ is constant. 
A Reynolds number for this problem can be defined in terms of the
heavy fluid density, the wavelength $W$ of the initial perturbation,
and the acceleration of gravity; that is, $Re=\rho_h\sqrt{W^3g}/\mu$,
which is set to 3000 for the 2D cases and 1024 for the 3D cases
reported here.

To ensure that the initial condition is consistent with conservation of
mass, the velocity field is prescribed as
\begin{equation}
\dfrac{\bm{u}}{\sqrt{Wg}} = -\dfrac{1}{Re\hspace{1pt}Sc} \dfrac{1}{\rho} \dfrac{d\rho}{dz} \grad{z}
\end{equation}

As described in \cref{subsec:bcs}, potential matching boundary
conditions are specified at the top and bottom boundary for the
fluctuating momentum, the mean momentum obeys a homogeneous Neumann
condition for the spanwise and streamwise components, and a
homogeneous Dirichlet boundary condition is specified for the vertical
component.  A homogeneous Neumann boundary condition is also used for
$z$.  See \cref{subsec:bcs} for more details on applying
boundary conditions in the algorithm presented here.

\begin{figure}[!h]
  \centering \includegraphics[]{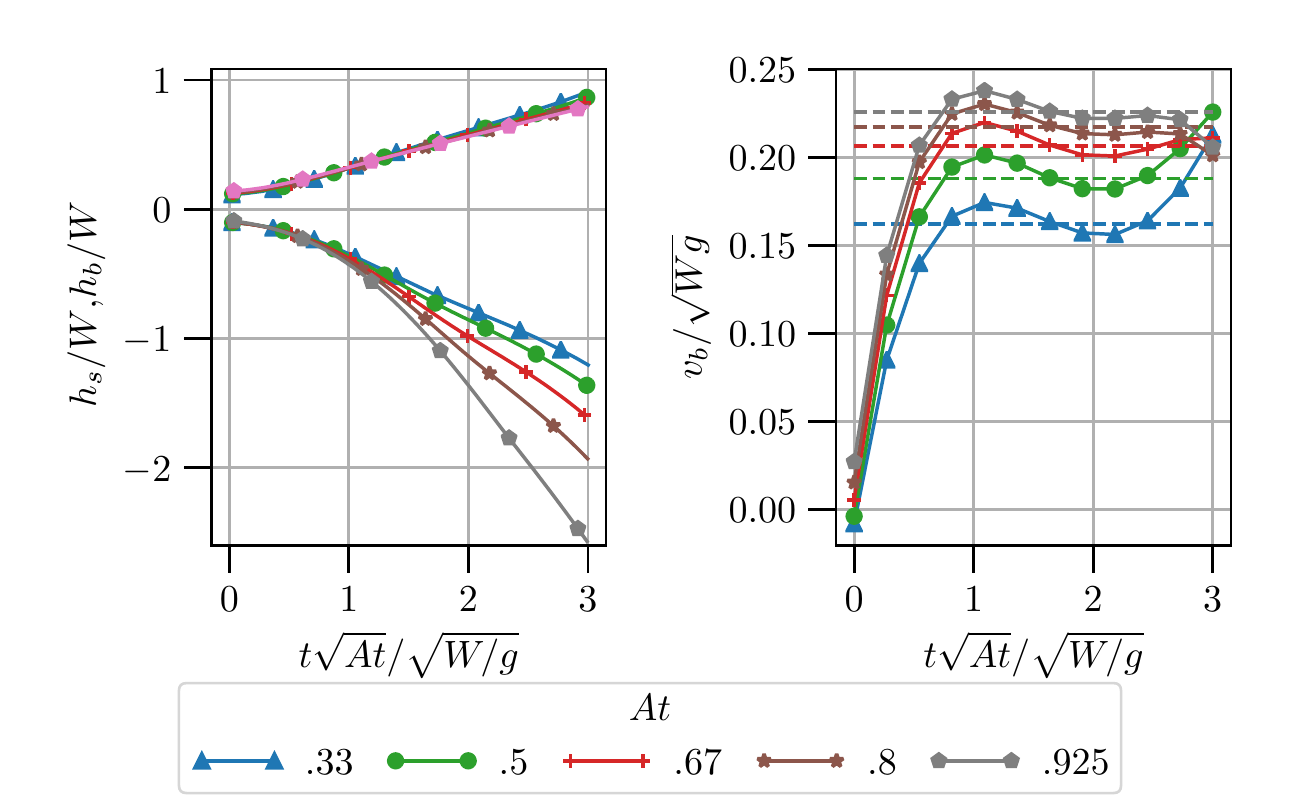} \caption[justification=raggedright]{\textit{Left}:
  Bubble and spike height ($h_b$,$h_s$) and \textit{right}: bubble
  velocity ($v_b$) as a function of time for various Atwood
  numbers. Two-dimensional cases with $Re = 3000$. Dashed horizontal
  lines are the velocity from potential theory, given by \cref{eqn:potential_growth}.} \label{fig:rt_2D_atsweep}
\end{figure}

The single-mode RT flow evolves through several different stages:
first, there is an initial acceleration when viscous effects dominate followed
by the formation and growth of a ``spike'' of heavy fluid traveling
downwards along with ``bubbles'' of light fluid penetrating upwards
into the heavy fluid (\cref{fig:rt_2D_cont_At_.5}).  Next, a period of
near constant bubble and spike velocity predicted by potential theory
occurs before a reacceleration (see, e.g., \cref{fig:rt_2D_atsweep}).
To verify the fidelity of the new algorithm, we compare simulation results to the
theoretical result of Goncharov \cite{Goncharov2002}, which gives the bubble
velocity, or the rate at which the light fluid penetrates the heavy,
during the potential growth phase as
\begin{equation}
v_b = \sqrt{\dfrac{2At}{1+At}\dfrac{g}{Ck}}
\label{eqn:potential_growth}
\end{equation}
where $C = 1$ in three-dimensional flows and $3$ in two-dimensional
flows and $k=2\pi/W$ is the perturbation wave number.  Additionally,
comparisons of the bubble/spike trajectories are made to other results
from the literature.

\begin{figure}[!h]
  \centering
  \includegraphics[]{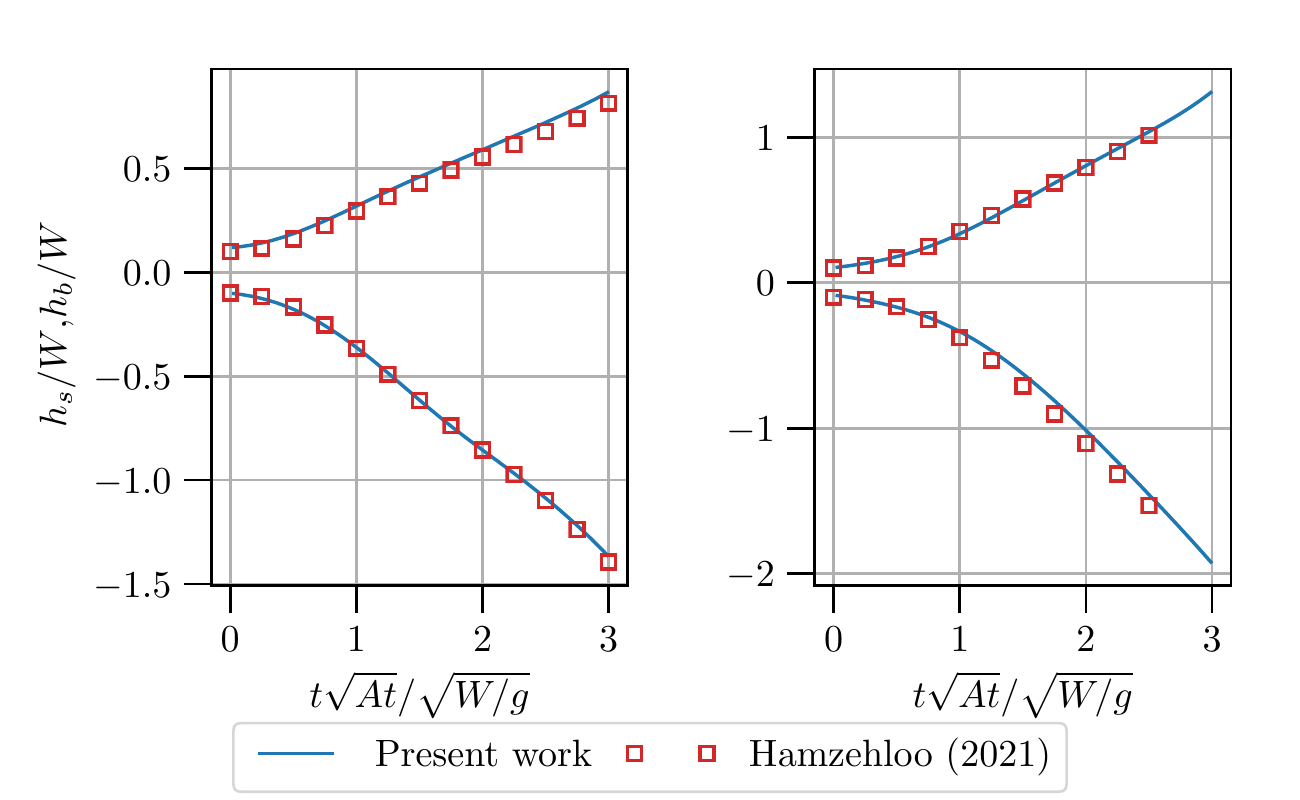}
  \caption[justification=raggedright]{Bubble and spike height
  ($h_b$,$h_s$) as a function of time -- $At = .5$. \textit{Left}: 2D
  case ($Re = 3000$), \textit{right}: 3D case ($Re = 1024$). Bubble velocities
  from Hamzehloo \cite{hamzehloo2021direct} also shown.} \label{fig:rt_2D3D_comp}
\end{figure}

To begin, we consider two validation cases found frequently in the
literature.  The first is a 2D case with $At = .5$ and $Re = 3000$ and
the second is a 3D case with $At = .5$ and $Re = 1024$. The evolution
of bubble and spike locations from the current calculations are in good
agreement with recent results from
Hamzehloo\etal\cite{hamzehloo2021direct}, as shown in
\cref{fig:rt_2D3D_comp}.
Furthermore, contours of density for the 2D case (\cref{fig:rt_2D_cont_At_.5}) are in good qualitative agreement with previous results \cite{hamzehloo2021direct,ding2007diffuse}.
As these previous results were for immiscible fluids using an interface-tracking or phase-field representation\todo{change or remove this detail?}, minor discrepancies are expected.

\begin{figure}[!h]
  \centering \includegraphics[]{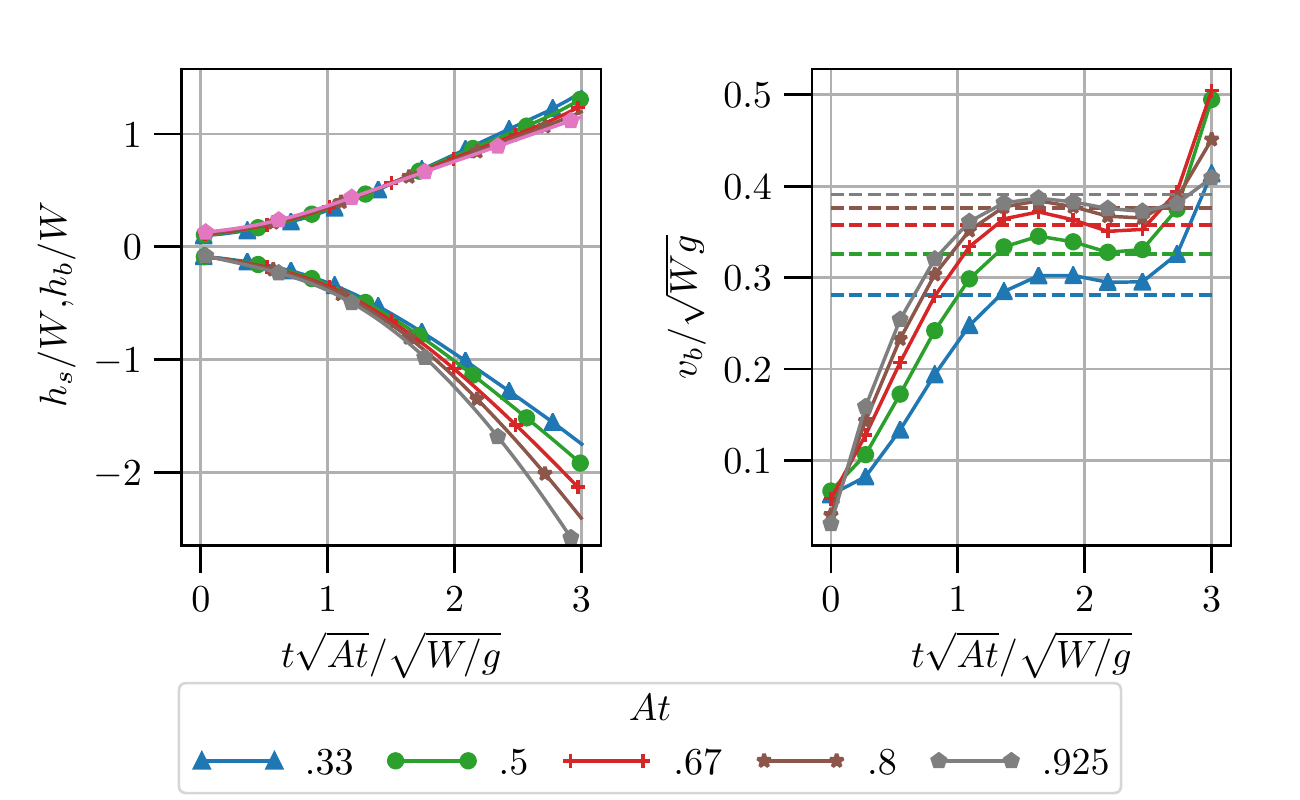} \caption[justification=raggedright]{\textit{Left}:
  Bubble and spike height ($h_b$,$h_s$) and \textit{right}: bubble
  velocity ($v_b$) as a function of time for various Atwood
  numbers. Three-dimensional cases with $Re = 1024$. Dashed horizontal
  lines are the velocity from potential theory, given
  by \cref{eqn:potential_growth}.} \label{fig:rt_3D_atsweep}
\end{figure}

Next, we explore the current algorithm's stability for increasing
density ratios by considering a range of Atwood number from 0.33 to
0.925, for both the
two- and three-dimensional flows. 
The maximum Atwood number corresponds to a density ratio of 25.67.
The bubble and spike evolution vary with Atwood number as expected
(\cref{fig:rt_2D_atsweep,fig:rt_3D_atsweep}).
%
%
Note that the bubble velocity during the potential growth phase, when the
velocity is approximately constant, is roughly consistent with the
potential theory \cref{eqn:potential_growth}. This consistency, along
with the consistency with \cite{hamzehloo2021direct} shown in \cref{fig:rt_2D3D_comp},
suggest that the algorithm is solving the variable-density equations
as intended. The minor discrepancies with potential theory can be
attributed to the more complete representation of the physics in the
variable-density equations\todo{this sentence got wonky}.  These results also
demonstrate the algorithm's robustness over a wide range of density
ratios.
\FloatBarrier

%% file: conclusion.tex
A new pseudospectral method has been developed for the direct
numerical simulation of low-Mach-number, variable-density flows that
are statistically homogeneous in two directions, allowing for the use
of Fourier representations in those directions. The method is designed
to preserve stability in high density ratio problems for explicit time
discretizations, which are attractive for DNS away from walls.  By
Helmoltz decomposing the momentum field, the momentum conservation
equations were recast to eliminate the pressure, as in
constant-density incompressible flows. The challenge was to treat the
redundancy between the mass conservation equation, the transport
equation for the scalar(s) that characterize the thermochemical state
of the fluid (e.g.~temperature, chemical species, or mixture fraction)
and the equation of state, which arise in the treatment of the
potential part of the momentum.  Analytically eliminating this
redundancy results in a more complex form of the scalar transport
equation, which, even in an explicit scheme, requires the solution of
a global linear system for the time derivative of the scalar that includes
the effect of convection by the velocity associated with the curl-free
momentum. An iterative solution scheme integrated with a novel
temporal discretization allows the target temporal order of accuracy
to be attained, independently of how well the linear solve is
converged. This allows the termination of the solution iterations
once the required stability of the temporal discretization is
attained.

The new method was tested in a series of single-mode Rayleigh-Taylor
instability problems. It was found to yield stable solutions for
density ratios up to about 26 that are both converged and consistent
with available theoretical and other computational results. This new
formulation greatly increases the density ratios that can be treated
without resorting to fully coupled implicit solvers \cite{Motheau2016,Knikker2011}, which due
to their expense are generally not appropriate for DNS of
turbulence.
\todo{RDM: this sentence OK? please add citations}
These may thus be useful for DNS of low-speed turbulent
combustion and low-speed turbulent mixing of fluids with different
densities. Combustion DNS is the application that motivated the
current developments.

There are several choices that were made in the formulation of the
method described here that may warrant further development. First,
the fixed point iteration in \cref{eqn:y_k} was used for its simplicity
in ensuring that any solution error was of the expected order in
time. It would be worthwhile exploring the use of more sophisticated
matrix-free solution algorithms. Also, the algorithm for determining
the potential momentum in \cref{subsec:CF_recon} is somewhat
complicated; a simpler and cheaper algorithm was also explored that
amounted to the solution of a Poisson equation for $\bm{m}^c$, but it
resulted in inferior stability for large density ratios. It would be
good to explore the possibility of gaining the advantages of the
latter scheme while preserving the stability advantages of the former.
Further, it may be possible to refine the time discretization and/or
the backward difference scheme used to obtain an initial guess for the
iterative solver to attain either better stability or higher accuracy,
and this would be worth pursuing. Finally, the formulation was presented
here for use with a single scalar transport equation, but in reacting
flow simulations there may be transport equations for many
thermochemical variables (e.g.~chemical species). The method needs to
be generalized for many scalars\todo{see note in tex file}.

%% file: coeffs.tex
The BDF-like approximations from \cref{sec:temp_disc} are generated by assuming the available time history of $z$ matches the progression of a variable time advanced by RK2.
That is,
\begin{equation}
\begin{aligned}
z^{n+1} &= z(t_n) + \dt \dzexact + \dfrac{1}{2} \dt^2 \ddzexact  + \mc{O}\left(\dt^3\right) \\
z' &= z(t_n) + \dfrac{1}{2} \dt \dzexact + \mc{O}\left(\dt^3\right) \\
z^n &= z(t_n) \\
z^{',n-1} &= z(t_n) - \dfrac{1}{2} \dt \dzexact + \mc{O}\left(\dt^3\right) \\
z^{n-1} &= z(t_n) - \dt \dzexact + \dfrac{1}{2} \dt^2 \ddzexact + \mc{O}\left(\dt^3\right) \\
z^{',n-2} &= z(t_n) - \dfrac{3}{2} \dt \dzexact + \dt^2 \ddzexact + \mc{O}\left(\dt^3\right) \, ,
\end{aligned}
\end{equation}
where Taylor expansions about $t_n$ have been used.
\subsection{Coefficients for fixed-point problem} \label{subsub:FP_coef}
\todo[inline]{Teresa's comment -- what do we do for the first few steps? Startup?}
We seek an approximation which is accurate up to $\mc{O}\left(\dt^2\right)$ to initialize the first stage fixed-point iteration (\cref{eqn:dz_fixedpoint_stageone}).
At this stage, $z^n, z^{n-1,'}, z^{n-1}, z^{n-2,'}$ are available.
Let
\begin{equation}
\begin{aligned}
\dt \dd{z}{t}^{n,0} &= \ai{1}^{n} z^n + \ai{2}^{n} z^{n-1,'} + \ai{3}^{n} z^{n-1} + \ai{4}^{n} z^{n-2,'} = \dt \dzexact + \mc{O}\left(\dt^3\right) \\
\end{aligned}
\end{equation}
then by matching the lower-order terms in the Taylor expansions, it follows that
\begin{equation}
\begin{aligned}
\ai{1}^{n} + \ai{2}^{n} + \ai{3}^{n} + \ai{4}^{n} &= 0 \\
-\dfrac{1}{2} \ai{2}^{n} - \ai{3}^{n} - \dfrac{3}{2} \ai{4}^{n} & = 1 \\
\dfrac{1}{2} \ai{3}^{n} + \ai{4}^{n} = 0 \, .
\end{aligned}
\end{equation}
This is satisfied by a one parameter family of solutions:
\begin{equation}
 \begin{alignedat}{4}
  &\ai{1}^{n} = 2 \quad &&\ai{2}^{n} = -2 - \dfrac{1}{2} \beta^n \quad && \ai{3}^n = \beta^n \quad &&\ai{4}^{n} = -\dfrac{1}{2}\beta^{n} \, .
 \end{alignedat}
\end{equation}

For the second stage, $z^{\prime},z^n, z^{n-1,'}, z^{n-1}$ are available for the approximation.
Similarly, letting
\begin{equation}
\begin{aligned}
\dt \dd{z}{t}^{',0} &= \ai{1}^{\prime} z^{\prime} + \ai{2}^{\prime} z^n + \ai{3}^{\prime} z^{n-1,'} + \ai{4}^{\prime} z^{n-1} \\ 
 &\quad\quad = \dt \dzexact + \dfrac{1}{2} \dt^2 \ddzexact + \mc{O}\left(\dt^3\right)
\end{aligned}
\end{equation}
and matching gives
\begin{equation}
\begin{aligned}
\ai{1}^{\prime} + \ai{2}^{\prime} + \ai{3}^{\prime} + \ai{4}^{\prime} &= 0 \\
\dfrac{1}{2} \ai{1}^{\prime} - \dfrac{1}{2} \ai{3}^{\prime} - \ai{4}^{\prime} &= 1 \\
\dfrac{1}{2} \ai{4}^{\prime} &= \dfrac{1}{2} \, .
\end{aligned}
\end{equation}
The family of solutions in terms of $\beta^{\prime}$ is
\begin{equation}
 \begin{alignedat}{4}
  &\ai{1}^{\prime} = 4 + \beta^{\prime} \quad &&\ai{2}^{\prime} = -5 - 2 \beta^{\prime} \quad && \ai{3}^{\prime} = \beta^{\prime} \quad &&\ai{4}^{\prime} = 1 \, .
 \end{alignedat} 
\end{equation}

%% file: lstab.tex
The linear stability analysis of \cref{subsec:linear_stab} is derived in terms of two matrices $\bm{A}_1$ and $\bm{A}_2$. 
These matrices are defined here.

Let $S = \sum_{j=0}^{k-1} \lambda_1^j  \lambda_2$. 
Then
\begin{equation}
\bm{A}_1= \begin{bmatrix}
1 + \lambda_1^k + \dfrac{1}{2} \dt S & -\lambda_1^k & 0 & 0 \\
0 & 1 & 0 & 0 \\ 0 & 0 & 1 & 0 \\ 0 & 0 & 0 & 1
\end{bmatrix}
\end{equation}
\begin{equation}
\bm{A}_2 = \begin{bmatrix}
1 + \left(3 + \beta'\right) \lambda_1^k + \dt S & - \left(6 + 2\beta'\right) \lambda_1^k - \dfrac{1}{2} \dt S & \beta' \lambda_1^k & \lambda_1^k \\
0 & 1 & 0 & 0 \\ 0 & 0 & 1 & 0 \\ 0 & 0 & 0 & 1
\end{bmatrix} \, .
\end{equation}

%% file: spat_disc.tex
In the representation of generic solution variables $f^h$
in \cref{eqn:disc_f}, the spatial representation in the $x_2$
direction remained unspecified because the algorithms presented here
are applicable to a variety of representations. However, some care is
required in the definition of the $x_2$ representation to ensure exact
mass conservation as described in \cref{subsec:CF_recon}. We begin by
defining a set of basis functions $\{B_j\}$ to represent the $x_2$
dependence of $\widehat{f}_{ln}$ in \cref{eqn:disc_f} that are smooth
enough to allow computation of the diffusion terms in the
Navier-Stokes and $z$ equations.\footnote{For the collocation method
used here in the $x_2$ direction, the $\{B_j\}$ need to be twice
differentiable.} Then \cref{eqn:disc_f} can be rewritten:
\begin{equation} \label{eqn:disc_f_inhomo}
\begin{aligned}
f^h(x_1,x_2,x_3,t) 
		   &=  \sum_{l = - \frac{N_1}{2} + 1}^{\frac{N_1}{2} - 1}
				 \sum_{j = 0}^{N_2 - 1}
				 \sum_{n = - \frac{N_3}{2} + 1}^{\frac{N_3}{2} - 1}
				 f_{ljn}(t) e^{\im k_1 x_1}
				 B_j(x_2)
				 e^{\im k_3 x_3} \\
&\quad k_1 = \dfrac{2\pi l}{L_1} \quad\quad k_3 = \dfrac{2\pi n}{L_3}
\end{aligned}
\end{equation}
This is the representation used for all the solution variables except
$m_2$. To ensure that mass conservation can be satisfied without
discretization error, we define a special set of basis functions
$\{B^\ast_j\}$ by which to represent $m_2$ and require that
$\left\{\frac{dB^\ast_j}{dx_2}\right\}$ and $\{B_j\}$
span the same space. In this way, in the mass conservation
equation \eqref{eqn:constraint1}, the time derivative of $\rho$ and all the terms in
the divergence of $\bm{m}$ will be in the same function space, so that the
equation can be satisfied exactly. For example, if $\{B_j\}$
are polynomials up to order $p$, then ${B^\ast_j}$ would be
polynomials up to order $p+1$. Similarly, if $\{B_j\}$ are piecewise
polynomials of order $p$ with $q$ continuous derivatives, than
$\{B^\ast_j\}$ would have to be piecewise polynomials of order $p+1$
with $q+1$ continuous derivatives.

\subsection{B-splines}
For the examples included in \cref{sec:results}, a maximum continuity
B-spline basis of order $p$ is used for all solution variables except
the $x_2$ component of momentum $m_2$, which is represented with
maximum continuity B-splines of order $p+1$.  B-splines of order $p$
are piecewise polynomials of degree $p-1$ defined by a set of knot
points $\left\{\xi_i\right\}$), which partition the domain into intervals.  The
B-splines have minimal local support and a maximum of $p-2$ continuous
derivatives at the knots.  They are attractive for the simulation of
turbulent
flows \cite{botella2003b,venugopal2008direct,Ulerich2014,lee2015direct,BAY2020109680}
due to their flexibility near boundaries and on nonuniform grids, and
their ability to achieve high spatial resolution while being
computationally efficient \cite{Kwok2001,botella2003b}.  A standard
reference on B-splines is de Boor \cite{DeBoor2001} and for specifics
about the use of B-splines with maximal continuity in fluid mechanics,
see \cite{botella2003b}.  With this solution representation, a
Fourier-Galerkin/B-spline-collocation method with approximate Galerkin
quadrature is used to obtain the spatially discrete
equations.

%% file: mesh_conv.tex
\begin{figure}[!h]
  \centering
  \includegraphics[]{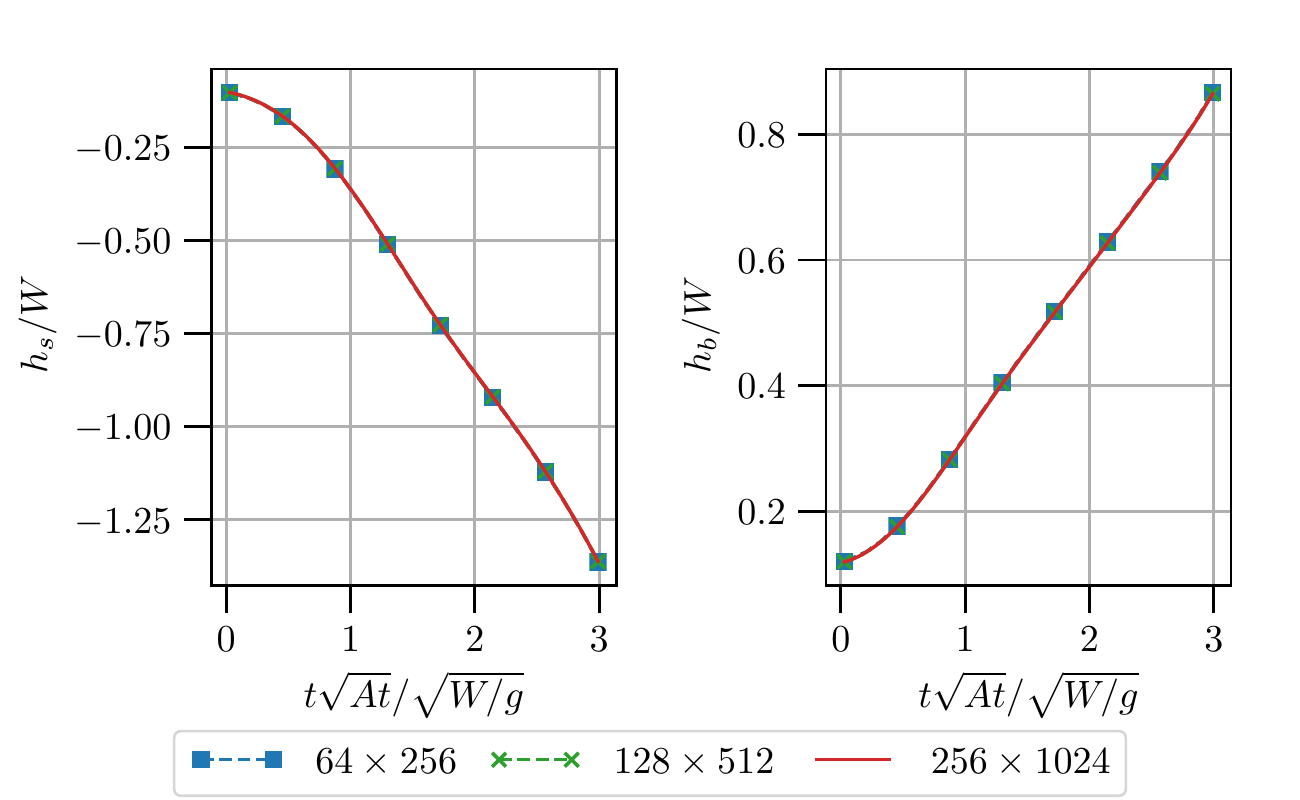}
  \caption[justification=raggedright]{Bubble and spike location as a
  function of time -- $At = .5$, 2-D RT problem. Results for three different meshes shown.} \label{fig:mesh_h}
\end{figure}
\begin{figure}[!h]
  \centering
  \includegraphics[]{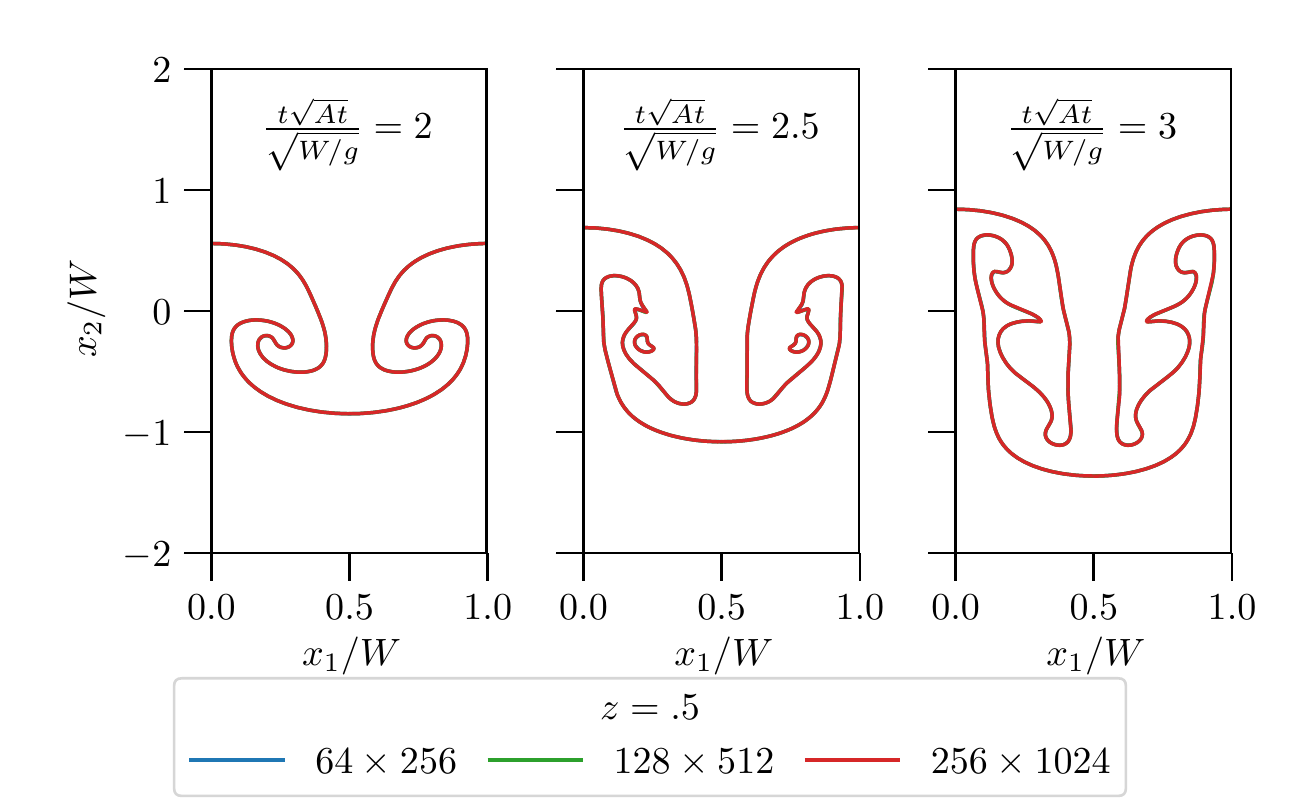}
  \caption[justification=raggedright]{Location and shape of the fluid interface (depicted by $z=.5$ contour) for late times -- $At = .5$, 2-D RT problem. Results for three different meshes shown.} \label{fig:mesh_z}
\end{figure}

To determine the spatial resolution used for the single-mode
Rayleigh-Taylor cases detailed in \cref{sec:results}, the 2D problem
with $At = .5$ and $Re = 3000$ was simulated with three different
meshes: $64 \times 256$, $128 \times 512$, and $256 \times
1024$.
Note that the Atwood number is low enough to use the smaller
computational domain in the vertical direction ($x_2\in[2W,2W]$).
The timestep details are given in \cref{tab:RT_casedetails}.
Comparisons of the bubble and spike
locations over time (\cref{fig:mesh_h}) as well as contours of the
interface for later times (\cref{fig:mesh_z}, $z = .5$) show no
discernible differences in the solutions with changing resolution.
The maximum variation of the bubble height $h_b/W$ amongst the different resolutions, for example, is $\sim .002$ at $t \sqrt{At}/\sqrt{W/g} = 3$ when the height itself is $\sim .8$.
A Richardson extrapolation procedure using the three solutions suggests an absolute error of $\mc{O}(10^{-4})$, corresponding to relative errors of $\mc{O}(10^{-2}) \%$, in the bubble height for late times.
As simulations for higher Atwood numbers were desired, the resolution suggested by the second grid was selected for the cases shown in this work.

For the cases presented in the text and this appendix, the computational domain, grid spacing, and timestep employed are listed in \cref{tab:RT_casedetails}.
The domains are defined as follows:
\begin{alignat*}{2}
\mc{V}_{RT}^{2D,1} &= [0, W] \times [-2W, 2W] \, , \qquad &128 \times 512 \, \text{mesh}; \\
\mc{V}_{RT}^{2D,2} &= [0, W] \times [-3.5W, 3.5W] \, , \qquad &128 \times 1024 \, \text{mesh}; \\
\mc{V}_{RT}^{3D,1} &= [0, W] \times [-2W, 2W] \times [0, W] \, , \qquad &128 \times 512 \times 128 \, \text{mesh}; \\
\mc{V}_{RT}^{3D,2} &= [0, W] \times [-3.5W, 3.5W] \times [0, W] \, , \qquad &128 \times 1024 \times 128 \, \text{mesh} \, .
\end{alignat*}
\renewcommand{\arraystretch}{1.5}
\begin{table}[]
\centering
\caption{Domain and timestep specification for Rayleigh-Taylor cases presented in \cref{sec:results}.}
\label{tab:RT_casedetails}
\begin{subtable}{.9\textwidth}
\centering
\begin{tabular}{ccccc}
\hline
$At$   & Dimension & $Re$ & Domain          & $\dt \sqrt{At}/\sqrt{W/g}$ \\ \hline
$.33$  & 2 & 3000 & $\mc{V}_{RT}^{2D,1}$ & $2.5 \times 10^{-4}$       \\
$.5$   & 2 & 3000 & $\mc{V}_{RT}^{2D,1}$ & $2.5 \times 10^{-4}$       \\
$.67$  & 2 & 3000 & $\mc{V}_{RT}^{2D,1}$ & $2.5 \times 10^{-4}$       \\
$.8$   & 2 & 3000 & $\mc{V}_{RT}^{2D,1}$ & $2.5 \times 10^{-4}$       \\
$.925$ & 2 & 3000 & $\mc{V}_{RT}^{2D,2}$ & $1.25 \times 10^{-4}$       \\ \hline
\end{tabular}
\end{subtable}
\\[10pt]
\begin{subtable}{.9\textwidth}
\centering
\begin{tabular}{ccccc}
\hline
$At$   & Dimension & $Re$ & Domain          & $\dt \sqrt{At}/\sqrt{W/g}$ \\ \hline
$.33$  & 3 & 1024 & $\mc{V}_{RT}^{3D,1}$ & $2.5 \times 10^{-4}$       \\
$.5$   & 3 & 1024 & $\mc{V}_{RT}^{3D,1}$ & $2.5 \times 10^{-4}$       \\
$.67$  & 3 & 1024 & $\mc{V}_{RT}^{3D,2}$ & $1.25  \times 10^{-4}$       \\
$.8$   & 3 & 1024 & $\mc{V}_{RT}^{3D,2}$ & $6.25 \times 10^{-5}$       \\
$.925$ & 3 & 1024 & $\mc{V}_{RT}^{2D,2}$ & $3.125 \times 10^{-5}$       \\ \hline
\end{tabular}
\end{subtable}
\end{table}